\documentclass[%
twocolumn,
superscriptaddress,
nofootinbib,
amsmath,amssymb,
aps,
prd,
]{revtex4-1}

\usepackage{graphicx}
\graphicspath{{.}{jripley_plots/}}
\usepackage{float}
\usepackage{bm}
\usepackage[hidelinks]{hyperref}
\usepackage{color}

\usepackage[utf8]{inputenc}
\usepackage[T1]{fontenc}
\usepackage{slashed}

\usepackage{fullpage}
\usepackage{enumerate}

\usepackage{mathalfa,amssymb,amsthm}
\usepackage{amsmath}

\begin{document}


\title{Evolution of Einstein-scalar-Gauss-Bonnet gravity 
   using a modified harmonic formulation}
\author{William E. East}
\email{weast@perimeterinstitute.ca}
\affiliation{%
Perimeter Institute for Theoretical Physics, Waterloo, Ontario N2L 2Y5, Canada.
}%
\author{Justin L. Ripley}
\email{lloydripley@gmail.com}
\affiliation{%
DAMTP,
Centre for Mathematical Sciences,
University of Cambridge,
Wilberforce Road, Cambridge CB3 0WA, UK.
}%

\date{\today}

\begin{abstract}
We present numerical solutions of several spacetimes of physical interest,
including binary black hole mergers,
in shift-symmetric Einstein-scalar-Gauss-Bonnet (ESGB)
gravity, and describe our methods for solving the full equations of motion,
without approximation, for general spacetimes.  While we concentrate on the
specific example of shift-symmetric ESGB, our methods, which make use of a
recently proposed modification to the generalized harmonic formulation, should
be generally applicable to all Horndeski theories of gravity
(including general relativity).  We demonstrate
that these methods can stably follow the formation of scalar clouds 
about initially vacuum non-spinning and spinning black holes
for values of the Gauss-Bonnet coupling
approaching the maximum value above which the hyperbolicity of the theory
breaks down in spherical symmetry.  We study the collision of
black holes with scalar hair, finding that the theory remains hyperbolic in the spacetime region
exterior to the black hole horizons in a similar regime, which includes cases
where the deviations from general relativity in the gravitational radiation is
appreciable.  Finally, we demonstrate that these methods can be used to follow
the inspiral and merger of binary black holes in full ESGB gravity.
This allows for making predictions for Horndeski theories of gravity
in the strong-field and non-perturbative regime,
which can confronted with gravitational wave observations,
and compared to approximate treatments of modifications to general 
relativity.

\end{abstract}

\maketitle

\section{Introduction}
\label{sec:introduction}
With the advent of gravitational wave astronomy, we are now in an unprecedented
position to test whether general relativity (GR) provides an accurate description of
gravity in the strong-field, highly dynamical regime. Observations of black
hole and other compact object mergers have already been used to constrain a
number of deformations of GR, including extra gravitational wave
polarizations, a graviton mass, and Lorentz violations~\cite{Abbott:2020jks}.
Despite these observational successes, for many alternative theories, it is
still unclear whether they are even on an equal
theoretical footing to GR, in the sense of being able to provide a full prediction of what
happens when two black holes merge.

Determining which theories that modify the principle part of the Einstein
equations are predictive in the strong-field regime (in the mathematical
sense that they have a well-posed initial value problem) has been a pressing
question in efforts to test such theories with observations.  Some theories,
for example dynamical
Chern-Simons~\cite{Delsate:2014hba,Okounkova:2018abo,Okounkova:2018pql}, or
theories that introduce Riemann-to-the-fourth-power type terms in the
action~\cite{Endlich:2017tqa,Cayuso:2020lca}, no longer have second order
equations of motion (EOMs), which is a requirement for a theory to be Ostrogradsky
stable \cite{Woodard:2015zca}. Thus one has little choice but to treat such
theories as valid on some limited range of scales, and to perturbatively
solve for the dynamics of those theories.
In contrast, here we will
concentrate on Einstein-scalar-Gauss Bonnet (ESGB), which is a representative example
of a Horndeski gravity theory.
Horndeski gravity theories are the class of
classical scalar-tensor theories that have second order EOMs
\cite{Horndeski:1974wa}, and thus can be thought of as the widest possible
class of scalar-tensor gravity theories that could act as classical,
theoretically viable alternatives to GR. 
Given this, Horndeski theories could be employed in
\emph{model dependent} tests of GR using gravitational wave
observations of black holes and neutron star binaries (for a recent review see,
e.g. \cite{Barack:2018yly}).   

As an alternative to thinking of modified gravity theories as 
complete classical field theories, the same as GR, one can 
view them as \emph{effective field theories}
(EFTs) that parameterize small deviations as a derivative expansion from GR (for
reviews see, e.g. \cite{Donoghue:1994dn,Burgess:2003jk}).  The dynamics of EFTs
are naturally solved in terms of an \emph{order reduction} approach, where
small corrections to Einstein evolution are solved for order by order in terms
of the effective coupling parameters.  
However, even taking this viewpoint,
there are reasons to study exact solutions to
Horndeski gravity theories.
While not all potential effective deviations from
GR are Horndeski theories (such as dynamical Chern-Simons gravity), a subclass
of Horndeski gravity theories called
``four derivative scalar-tensor theory'' (4$\partial$ST)
gravity encompasses the leading
order scalar-tensor interactions that are parity invariant
\cite{Weinberg:2008hq}.  One challenge with computing dynamical solutions to
EFTs through an order reduction approach is that the solution can be
contaminated by secular effects, which are purely artifacts of the order
reduction approximation, but which can grow in time.  There have been some
proposals to address these problems~\cite{Okounkova:2017yby,Okounkova:2020rqw},
including by changing the behavior of the theory at short wavelengths in an
ad-hoc manner in order to cure problems with
well-posedness~\cite{Cayuso:2017iqc}.  However, without full
solutions to compare to, it is difficult to quantify the errors introduced by
these methods.  Stable numerical solutions to the exact EOMs 
would not be subject to those kinds of secular effects. For small enough
modified gravity couplings, exact solutions could be used to find essentially
perturbative corrections to the Einstein equations, while avoiding spurious
contamination of the solution from secular errors.  

One technical challenge that has prevented finding fully nonlinear solutions to
many Horndeski gravity theories for spacetimes that lack any symmetries (as is
the case for binary black hole merger spacetimes), has been that there was no
known well-posed initial formulation for Horndeski gravity.  Recently though,
Kovacs and Reall proved that the equations of motion for Horndeski gravity
theories possess a well posed initial value problem in a \emph{modified
harmonic formulation} \cite{Kovacs:2020pns,Kovacs:2020ywu}\footnote{We note
that a different gauge condition was proven to provide a well-posed initial
problem for ``cubic'' Horndeski theories (which do not include ESGB) by Kovacs \cite{Kovacs:2019jqj}, and
that gauge condition was numerically implemented in 
Ref.~\cite{Figueras:2020dzx} to study spherical collapse in that class
of theories.}, as long as the coupling parameter that determines the
beyond-GR corrections is much smaller than all other length scales in the
problem.  Their result has opened up the possibility for full numerical
simulations of Horndeski gravity theories for spacetimes of physical interest,
such as binary black hole spacetimes, and for cosmologies that are not
perfectly homogeneous. However, determining for what range of couplings this
formulation is hyperbolic for strong-field, dynamical spacetimes is something
that most likely needs to be done on case-by-case basis numerically, for
different Horndeski theories and choices of initial data.
   
Here, we numerically solve for the dynamics of black holes
in \emph{shift-symmetric} ESGB gravity.
Shift-symmetric ESGB gravity has attracted recent interest as the 
Schwarzschild and Kerr black hole solutions of GR are not stationary
solutions to this theory \cite{Sotiriou:2014pfa}: vacuum black holes
will evolve to black hole solutions with 
stable scalar field clouds (i.e. scalar hair) around them. 
Disregarding Horndeski models that model dark
energy (many of which have been highly constrained, see e.g.
\cite{Arai:2017hxj,Ezquiaga:2017ekz,Kase:2018aps}), ESGB 
gravity have attracted recent attention
as the couplings for the theory are relatively weakly constrained, yet the
theory admits scalar hairy black hole solutions (see e.g.
\cite{Kanti:1997br,Sotiriou:2013qea}), the collision of which should produce
gravitational wave signals that differ noticeably from those of GR black holes.
ESGB gravity thus promises to act as a useful foil to perform model
dependent tests of GR in the strong field, dynamical regime (for further
discussion see, e.g., \cite{Berti:2018cxi}).

Shift symmetric ESGB gravity can be motivated as the leading order scalar
tensor theory of gravity whose equations of motion are invariant
under shifts in the scalar field: $\phi\to\phi+const.$
\cite{Weinberg:2008hq,Sotiriou:2013qea}\footnote{Although in
this paper, for simplicity we do not
consider \emph{all} terms allowed by shift symmetry; in particular we
do not consider the term $\alpha X^2$;
see Eq.~\eqref{eq:leading_order_in_derivatives_action}.}.
While there is an EFT  argument to motivate this theory,
we solve for the full equations of motion
without any perturbative assumptions, potentially
outside the regime of validity of the assumptions of EFT.
While in this article we only consider numerical solutions to
shift-symmetric ESGB gravity, for general interest we also present the
equations of motion for the leading order scalar tensor theory
of gravity whose equations of motion are invariant under parity
inversion: $x^a\to-x^a$ (4$\partial$ST gravity),
in a form suitable for use in modified harmonic evolution. 

   Earlier studies of shift-symmetric ESGB gravity have been limited to either
spacetimes with a high degree of symmetry (e.g.
\cite{Sotiriou:2014pfa,Ripley:2019irj,Ripley:2019aqj}),
stationary solutions~\cite{Delgado:2020rev,Sullivan:2020zpf}, or 
to perturbative/order reduction solutions to the theory
(e.g. \cite{Benkel:2016rlz,Benkel:2016kcq,Witek:2018dmd,Okounkova:2019zep,Okounkova:2020rqw}).
In Refs.~\cite{Ripley:2019irj,Ripley:2019aqj}, it was found that
in spherical symmetry, for sufficiently large values of the Gauss-Bonnet coupling, black hole 
spacetimes could develop elliptic regions, where the hyperbolicity of the
equations broke down, outside the black hole horizon.
This sets an upper bound for the range of parameters where the
theory will remain hyperbolic once spherical symmetry is broken.

In this paper, we describe our methods for numerically solving the full
equations of ESGB gravity and use them to study the dynamics
of black hole scalar hair formation and black hole mergers.
One of our main results is that we find that we are able
to solve for spacetimes where the deviations from GR are significant in
terms of the changes to the black holes due to dynamical scalar hair formation, 
and the imprint on the gravitational waves. 
The remainder of the paper is as follows. In Sec.~\ref{sec:eom_mod_harm}, 
we present the EOMs for the general class of 4$\partial$ST
gravity (which includes ESGB)
in the form we use for numerical evolutions. In Sec.~\ref{sec:numerical}, we describe
our numerical methods for evolving these equations. 
In Sec.~\ref{sec:results}, we presents our results, beginning with 
a robustness test to illustrate the improved hyperbolicity, and 
then moving on to several physically interesting problems
including the dynamical formation of scalar hair about
spinning black holes in axisymmetry and a fully 3D setting, and
head-on and quasi-circular binary black hole mergers.
We discuss these results and conclude in Sec.~\ref{sec:conclusion}.

In this work we use geometric units: $G=c=1$,
a metric sign convention of $-+++$,
lower case Latin letters to
index spacetime indices, and lower case Greek letters to index spatial
indices.
\section{Equations of motion}

\subsection{Modified generalized harmonic formulation}
\label{sec:eom_mod_harm}
We begin by briefly reviewing
the modified generalized harmonic (MGH) formulation
\cite{Kovacs:2020pns,Kovacs:2020ywu}. 
In a Lorentzian spacetime $(M,g)$, we introduce two auxiliary
Lorentzian metrics $\tilde{g}^{mn}$ and $\hat{g}^{mn}$.
We will always raise and lower indices with
the spacetime metric $g_{ab}$,
so e.g. $\hat{g}^{ab}\equiv g^{ac}g^{bd}\hat{g}_{cd}$.
We also define $\tilde{g}\equiv \tilde{g}^{ab}g_{ab}$ and
$\hat{g}\equiv\hat{g}^{ab}g_{ab}$.
The MGH formulation imposes the following
conditions on the coordinates $x^{c}$:
\begin{align}
\label{eq:mh_condition}
    C^{c}
    \equiv&
    H^c
-	\tilde{g}^{ab}
    \nabla_{a}\nabla_{b}x^{c}
    \nonumber\\
    =&
    H^c
+	\tilde{g}^{ab}\Gamma_{ab}^c
    =
    0
    .
\end{align} 
    As in the generalized harmonic formulation,
$H^c$ are the source functions that, along with $\tilde{g}^{ab}$,
determine the gauge degrees of freedom, and
$C^c$ (which will generally not be exactly zero in a given numerical solution)
is called the \emph{constraint violation}.
We next define the MGH EOMs as 
\begin{align}
\label{eq:basic_eom}
    &E^{ab}
-	\hat{P}_{d}{}^{cab}
    \nabla_{c}C^{d}
\nonumber\\
&- \frac{1}{2}\kappa\left(
            n^aC^b
    +	n^bC^a
    +	\rho n^cC_c g^{ab}
    \right)
    =
    0
    ,
\end{align}
    where $E^{ab}$ are the EOMs derived from varying the
    metric,
$n^a$ is a time-like vector (we assume
$n^a$ is timelike with respect to $g^{ab}$,
$\tilde{g}^{ab}$, and $\hat{g}^{ab}$), and
\begin{align}
    \hat{P}_{d}{}^{cab}
    \equiv
    \frac{1}{2}\left(
            \delta_d^a\hat{g}^{bc}
    +	\delta_d^b\hat{g}^{ac}
    -	\delta_{d}^{c}\hat{g}^{ab}
    \right)
    .
\end{align}
We include constraint damping with the constants $\kappa$ and $\rho$
\cite{Gundlach:2005eh}
\footnote{Note that we need $\kappa<0$ to damp out the constraints. Also, some
of our sign conventions differ from ~\cite{Kovacs:2020ywu}.
}. 
Note as well that Eq.~\ref{eq:basic_eom} is slightly different
from Ref.~\cite{Kovacs:2020ywu}: here we use $\nabla_{c}C^{d}$ instead of
$\partial_{c}C^{d}$.  We choose the form used here for consistency with
the standard generalized harmonic formulation (see
Appendix~\ref{eq:reduction_eom_Einstein}), though either way the principal
part, and hence the hyperbolicity results, will be the same.
   From Eq.~\ref{eq:mh_condition},
we see that in the MGH formulation the coordinates
$x^a$ obey a hyperbolic equation with characteristics determined by
$\tilde{g}^{ab}$. Taking the divergence of Eq.~\ref{eq:basic_eom},
and assuming $\nabla_aE^{ab}=0$ (which holds for all the theories
we consider, including the Einstein equations),
we obtain a hyperbolic equation
for the constraint violating modes $C^a$: 
\begin{align}
\label{eq:eom_constraint_violating}
-  \frac{1}{2}\hat{g}^{ac}\nabla_a\nabla_cC^b
-  \hat{g}^{cb}R_{dc}C^d
-  \left(\nabla_a\hat{P}_d{}^{cab}\right)\left(\nabla_cC^d\right)
\nonumber \\
-  \frac{1}{2}\kappa\nabla_a\left(
      n^aC^b
   +  n^bC^a
   +  \rho n^cC_c g^{ab}
   \right)
   =0
   .
\end{align}
From Eq.~\ref{eq:eom_constraint_violating},
we see that the constraint violating modes 
obey a hyperbolic equation with characteristics determined
by $\hat{g}^{ab}$.
With the special choice of $\tilde{g}^{ab}=\hat{g}^{ab}=g^{ab}$, the MGH 
formulation reduces to the generalized harmonic formulation
(for an explicit calculation of this in the context of the
Einstein equations, see Appendix~\ref{eq:reduction_eom_Einstein}).
Finally, we note that
picking a gauge in the MGH formulation amounts to choosing the
functional form of the auxiliary metrics $\tilde{g}^{ab}$ and $\hat{g}^{ab}$,
and choosing the functional form of the source function $H^c$.
\subsection{Equations of
four derivative scalar tensor (4$\partial$ST) gravity
for a modified harmonic formulation}
\label{sec:eom_derivation}

   While we only consider numerical solutions to shift-symmetric ESGB
gravity in this article,
we derive the EOMs in the MGH formulation
for the following scalar-tensor theory
(Kovacs and Reall call 4$\partial$ST gravity
\cite{Kovacs:2020pns,Kovacs:2020ywu}),
for which ESGB gravity is a specific example.
We do this for the sake of generality, and given the applications of
4$\partial$ST gravity in, e.g., EFTs of the early universe
(e.g. \cite{Weinberg:2008hq}).
The action is:
\begin{align}
\label{eq:leading_order_in_derivatives_action}
    S
    =&
    \frac{1}{8\pi}\int d^4x\sqrt{-g}
 \nonumber \\ 
&\left(
        \frac{1}{2}R
    +	X
    -	V\left(\phi\right)
    +	\alpha\left(\phi\right)X^2
    +	\beta\left(\phi\right)\mathcal{G}
    \right)
    ,
\end{align}
    where
\begin{subequations}
\begin{align}
    X
    \equiv&
-	\frac{1}{2}\left(\nabla\phi\right)^2
    ,\\
    \mathcal{G}
    \equiv&
    \frac{1}{4}
    \delta^{abcd}_{efgh}R^{ef}{}_{ab}R^{gh}{}_{cd}
    , \\
    \delta^{abcd}_{efgh}
    \equiv&
    4!\delta^a_{[e}\delta^b_f\delta^c_g\delta^d_{h]}
    ,
\end{align}
\end{subequations}
   and $V$, $\alpha$, and $\beta$ are functions of $\phi$.
If one interprets Eq.~\ref{eq:leading_order_in_derivatives_action}
as an EFT, it contains (up to total derivatives,
field redefinitions, and conformal rescalings)
all scalar-tensor terms involving up to four derivatives.
Thus, from an EFT perspective, the theory
represents the leading order (in derivatives) scalar-tensor theory 
that is preserved under parity transformations
\cite{Weinberg:2008hq,Kovacs:2020pns,Kovacs:2020ywu}\footnote{
   For more context regarding the theory we consider:
Ref.~\cite{Weinberg:2008hq} considered the Weyl tensor coupling
$f(\phi)C_{abcd}C^{abcd}$, 
which when varied in the action leads to fourth order equations
of motion, which likely do not have a well-posed initial value formulation
when taken as classical PDE.
Refs.~\cite{Kovacs:2020pns,Kovacs:2020ywu}
pointed out that if (through field redefinitions) one replaces 
$C_{abcd}C^{abcd}$ with the Gauss-Bonnet
scalar $\mathcal{G}$, then the EOMs
are second order in time and space,
and furthermore have a well-posed initial
value problem in the MGH formulation.
}.

   We obtain shift-symmetric ESGB gravity by choosing $V(\phi)=\alpha(\phi)=0$
   and $\beta(\phi)=\lambda\phi$.
Here $\lambda$ is a constant coupling parameter, that in
geometric units has dimensions of length squared.
As the Gauss-Bonnet scalar $\mathcal{G}$ is a total derivative
in four dimensions, we see that the action of shift-symmetric ESGB gravity is preserved
up to total derivatives under constant shifts in the scalar field:
$\phi\to\phi+\textrm{constant}$.

   Varying Eq.~\ref{eq:leading_order_in_derivatives_action} with
respect to the scalar field and metric
gives us the EOMs 
\begin{align}
\label{eq:eom_edgb_scalar}
   E^{(\phi)}
   &\equiv
      \Box\phi
   -  V^{\prime}\left(\phi\right)
   \nonumber\\&
   +  2\alpha\left(\phi\right)X \Box\phi
   -  2\alpha\left(\phi\right)\nabla^a\phi\nabla^b\phi\nabla_a\nabla_b\phi
   \nonumber\\&
   -  3\alpha^{\prime}\left(\phi\right)X^2
   +  \beta^{\prime}\left(\phi\right)\mathcal{G}
   =
   0
   ,\\
\label{eq:eom_edgb_tensor}
   E^{(g)}_{ab}
   &\equiv
   R_{ab}
-  \frac{1}{2}g_{ab}R
   \nonumber\\&
-  \nabla_a\phi\nabla_b\phi
+  \left(-X+V\left(\phi\right)\right)g_{ab}
   \nonumber\\&
-  2\alpha\left(\phi\right)X\nabla_a\phi\nabla_b\phi
-  \alpha\left(\phi\right)X^2g_{ab}
   \nonumber\\&
+  2\delta^{efcd}_{ijg(a}g_{b)d}R^{ij}{}_{ef}
   \nabla^g\nabla_c\beta\left(\phi\right) 
   =
   0
   .
\end{align}
   We take the trace-reverse of Eq.~\ref{eq:eom_edgb_tensor} to obtain 
\begin{align}
\label{eq:eom_edgb_tensor_trace_reversed}
   E^{(g,TR)}_{ab}
   \equiv &
   E^{(g)}_{ab}-\frac{1}{2}g_{ab}E^{(g)}
    \nonumber\\
    =&
    R_{ab}
-   \nabla_a\phi\nabla_b\phi
-   V\left(\phi\right)g_{ab}
    \nonumber\\&
-   2\alpha\left(\phi\right)X\nabla_a\phi\nabla_b\phi
-   \alpha\left(\phi\right)X^2g_{ab}
    \nonumber\\&
+   2\delta^{efcd}_{ijg(a}g_{b)d}R^{ij}{}_{ef}
      \nabla^g\nabla_c\beta\left(\phi\right)
    \nonumber\\&
-   \delta^{efc}_{ijg}R^{ij}{}_{ef}\nabla^g\nabla_c\beta\left(\phi\right)
    g_{ab}
    .
\end{align}
   The last step we take before expanding out the EOMs 
is to add in the MGH constraint propagation term and
a constraint damping term \cite{Gundlach:2005eh,Kovacs:2020ywu}:
\begin{align}
\label{eq:eom_edgb_damped}
   E^{(g,C)}_{ab}
   \equiv &
   E^{(g,TR)}_{ab}
-  \left(
      \hat{P}_c{}^d{}_{ab}
   -  \frac{1}{2}g_{ab}\hat{P}_c{}^d
   \right)
   \nabla_dC^c
   \nonumber\\&
-  \frac{1}{2}\kappa\left(
      n_aC_b
   +  n_bC_a
   -  \left(1+\rho\right) n_cC^cg_{ab}
   \right)
   ,
\end{align}
   where we have defined $\hat{P}_d{}^c\equiv g_{ab}\hat{P}_d{}^{cab}$.

   Finally, we rewrite the 4$\partial$ST EOMs,
Eqs.~\ref{eq:eom_edgb_tensor_trace_reversed} and \ref{eq:eom_edgb_damped},
in the following form
\begin{align}
\label{eq:time_evo_system}
   \begin{pmatrix}
      A_{ab}{}^{cd}
   &  B_{ab}
   \\ C^{cd}
   &  D
   \end{pmatrix}
   \partial_0^2
   \begin{pmatrix}
      g_{cd}
   \\ \phi
   \end{pmatrix}
+  \begin{pmatrix}
      F_{ab}^{(g)}
   \\ F^{(\phi)}
   \end{pmatrix}
   =
   0
   .
\end{align}
   In Appendix~\ref{appen:derivation_evolution_matrix}, we derive
the explicit forms of the components $A_{ab}{}^{cd}$, etc. for the
4$\partial$ST EOMs.
For the remainder of this paper we will restrict our
attention to the particular case of shift-symmetric ESGB gravity.
\section{Numerical implementation}
\label{sec:numerical}
In this section, we describe our methods for numerically evolving the
equations of shift-symmetric ESGB gravity, which we recall is a special
case of 4$\partial$ST gravity with $V(\phi)=\alpha=0$ and $\beta=\lambda\phi$.
Our general strategy is to, 
where possible, adapt the methods of
Ref.~\cite{Pretorius:2004jg} for evolving Einstein gravity in
a generalized harmonic formulation to these new equations.


\subsection{Form of equations of motion and gauge choices}
We directly evolve the 22 variables (after accounting for the symmetry
of the metric components)
$\{g_{ab},\ \partial_0 g_{ab},\ \phi, \partial_0 \phi\}$
using the EOM given by Eq.~\ref{eq:time_evo_system}. 

In addition to the physical metric, 
we must also specify the auxiliary metrics $\tilde{g}^{ab}$ and $\hat{g}^{ab}$. 
In general, there is a large degree of freedom in choosing these as functions
of $g_{ab}$ and the spacetime coordinates, though
here we restrict to a relatively simple choice given by
\begin{subequations}
\label{eq:relation_gtilde_ghat_g}
\begin{align}
   \tilde{g}^{ab}
   = &
   g^{ab} - \tilde{A} n^{a}n^{b}
   ,\\
   \hat{g}^{ab}
   = &
   g^{ab} - \hat{A} n^{a}n^{b}
   ,
\end{align}
\end{subequations}
   where $n^{a}$ is the (time-like) unit normal vector to the spacelike
hypersurfaces we evolve on, and $\tilde{A}$ and $\hat{A}$ are constants.
We emphasize that the MGH formulation only requires $\tilde{g}^{ab}$
and $\hat{g}^{ab}$ to be Lorentzian. We have chosen the 
ansatz \ref{eq:relation_gtilde_ghat_g} out of its simplicity to implement,
and empirically we find that we are able to numerically solve
the ESGB equations of motion using auxiliary metrics of this form.

As in the generalized harmonic formulation, we must also choose the source functions $H_a$,
which, combined with the auxiliary metric, determine the coordinate degrees of freedom.
Here we restrict to the damped harmonic gauge~\cite{Lindblom:2009tu,Choptuik:2009ww}
(including the special case of $H_a=0$), 
which has been found to work well for a large number of highly dynamical spacetimes,
or fix $H_a$ to be constant in time for some cases
where we wish to maintain Kerr-Schild like coordinates.

\subsection{Numerical discretization}
The numerical scheme we use follows that of Ref.~\cite{East:2011aa}.
We discretize the partial differential equations in space, using standard
fourth-order finite difference stencils, and in time, using fourth-order
Runge-Kutta integration.
We implement the EOM directly in the form given by Eq.~\ref{eq:time_evo_system},
and invert the set of linear equations at each point using Gaussian elimination.
We control high frequency numerical
noise using Kreiss-Oliger dissipation~\cite{1972Tell...24..199K}.
As indicated in Eq.~\ref{eq:basic_eom}, 
we also use constraint damping to control the constraint
violating modes sourced by truncation error. We typically set $\kappa=-1/M_{\rm BH}$, where $M_{\rm BH}$
is the mass of the smallest black hole in the simulation, and $\rho=0$,
which are similar values to those used
in black hole evolutions using the generalized harmonic formulation~\cite{Pretorius:2006tp}.

As detailed in Ref.~\cite{Pretorius:2004jg},
we use compactified coordinates so that physical 
boundary conditions
(namely that the metric is flat and the scalar field vanishes)
can be placed at spatial infinity.
We use Berger-Oliger~\cite{1984JCoPh..53..484B} style adaptive mesh refinement
(AMR) supported by the PAMR library~\cite{Pretorius:2005ua,PAMR_online}.
The interpolation in time for the AMR boundaries is only third-order accurate,
which can reduce the overall convergence to this order in some instances. 
In some of the cases here, we restrict to axisymmetric spacetimes,
and use the modified Cartoon method
to reduce our computational domain to a
two-dimensional Cartesian half-plane~\cite{Pretorius:2004jg}.

\subsection{Excision}
A crucial ingredient in our ability to evolve black hole spacetimes is the use of
excision. In ESGB, the situation is worse than in GR since, as shown in
Ref.~\cite{Ripley:2019irj}, elliptic regions can develop just inside a black hole
horizon, where the EOMs are no longer well-posed, despite the region having
bounded curvature.  Following Ref.~\cite{Pretorius:2004jg}, we dynamically
track any apparent horizons in our spacetime and excise an interior region.
This is done by finding an ellipsoid that just fits inside the apparent horizon
and shrinking the axes, typically by 15 to $25\%$, to create a buffer region
between the apparent horizon and the excision surface.  In general, we find
that we must use smaller excision buffers as the coefficient of the modified
gravity terms (i.e. $\lambda$) is increased, which requires higher resolution,
in order to avoid instabilities near the excision surface.

As the apparent horizon evolves, points that were previously excised may become
unexcised and need to be ``repopulated" by extrapolating their values from
neighboring points.  When evolving with unmodified GR equations, this is often
done with simple first-order extrapolation, i.e. by taking the average value of
the neighboring unexcised points, to avoid high frequency noise (and since the
points should initially be out of causal contact with the exterior domain).
However, we find that when evolving with ESGB we must use second-order or higher
extrapolation, which we speculate is due to the presence of terms of the form
$(\partial \partial g)^2$ and $(\partial\partial g)(\partial \partial \phi)$ in
the EOM, which are sensitive to jumps in the second derivative.
We note that a possible alternative to the excision method used here is to
modify the EOMs inside black hole horizons---e.g. by letting the non-GR
coupling go to zero---so that they remain hyperbolic~\cite{Figueras:2020dzx}. 
\subsection{Initial data}
\label{sec:initial_data}
   On our initial data surface,
we must satisfy the generalizations of the Hamiltonian constraint
$\mathcal{H}\equiv n^an^b\mathcal{E}_{ab}^{(g)}$
and momentum constraint $\mathcal{M}_{\gamma}\equiv n^aE^{(g)}_{a \gamma}$,
which for 4$\partial$ST gravity take the form
\begin{subequations}
\label{eq:constraints}
\begin{align}
\label{eq:hamiltonian_constraint}
   \mathcal{H}
   =&
   n^an^bR_{ab}
+  \frac{1}{2}R
   \nonumber\\&
-  \left(n^a\nabla_a\phi\right)^2
+  X - V\left(\phi\right)
   \nonumber\\&
-  2\alpha\left(\phi\right)X\left(n^a\nabla_a\phi\right)^2
+  \alpha\left(\phi\right)X^2
   \nonumber\\&
+  2n^an^b\delta^{efcd}_{ijg(a}g_{b)d}
   R^{ij}{}_{ef}\nabla^g\nabla_c\beta\left(\phi\right)
   \\
\label{eq:momentum_constraint}
   \mathcal{M}_{\gamma}
   =&
   n^aR_{a\gamma}
   \nonumber\\&
-  n^a\nabla_a\phi\nabla_{\gamma}\phi
   \nonumber\\&
-  2\alpha\left(\phi\right)Xn^a\nabla_a\phi\nabla_{\gamma}\phi
   \nonumber\\&
+  2n^a\delta^{efcd}_{ijg(a}g_{\gamma)d}
   R^{ij}{}_{ef}\nabla^g\nabla_c\beta\left(\phi\right)
   .
\end{align}
\end{subequations}
   Here, we do not implement a method to solve these equations 
for general $\phi$. Instead, we consider
initial data for which $\phi=\partial_0\phi=0$ on the initial data surface.
With this choice of $\phi$
the ESGB contributions to the constraint equations, which we
define to be:
\begin{subequations}
\label{eq:edgb_contribution_constraints}
\begin{align}
   \mathcal{H}^{(GB)}
   \equiv&
   2 n^an^b\delta^{efcd}_{ijg(a}g_{b)d}
   R^{ij}{}_{ef} 
      g^{gk}\partial_k\partial_c\beta\left(\phi\right)
   \\
   \mathcal{M}^{(GB)}_{\gamma}
   \equiv&
   2n^a\delta^{efcd}_{ijg(a}g_{\gamma)d}
   R^{ij}{}_{ef}
      g^{gk}\partial_k\partial_c\beta\left(\phi\right)
   ,
\end{align}
\end{subequations}
   are identically zero on the initial data surface. 
To show this, we first expand
Eqs.~\ref{eq:edgb_contribution_constraints},
imposing $\phi=\partial_0\phi=0$ (which 
implies, e.g. $\partial_{\alpha}\partial_0\phi=0$),
rewriting terms to include the unit normal 
to slices of constant time: $n_a=\left(-N,0,0,0\right)$
(here $N=1/\sqrt{-g^{tt}}$ is the lapse function)
so that we are left with
\begin{subequations}
\label{eq:edgb_contribution_constraints_expanded}
\begin{align}
   &\mathcal{H}^{(GB)}
   =
   \nonumber\\
   &2 \left(-\frac{1}{N}\right) n^an^bn^gn_q\delta^{efqd}_{ijg(a}g_{b)d}
   R^{ij}{}_{ef}
      \beta^{\prime}\left(\phi\right)
      \partial_0^2\phi
   ,\\
   &\mathcal{M}^{(GB)}_{\gamma}
   =
   \nonumber\\
   &2\left(-\frac{1}{N}\right)n^an^gn_q\delta^{efqd}_{ijg(a}g_{\gamma)d}
   R^{ij}{}_{ef}
      \beta^{\prime}\left(\phi\right)
      \partial_0^2\phi
   .
\end{align}
\end{subequations}
   We see that the $n^a$ vectors
symmetrize the totally antisymmetric indices of the
generalized Kronecker delta; e.g. $n^an^g\delta^{efcd}_{ijga}=0$,
so that the ESGB contributions to the constraints
equations on the initial data surface vanish.

   Thus, in shift-symmetric ESGB gravity ($\alpha=V=0$, $\beta=\lambda\phi$),
and with scalar field initial data $\phi=\partial_0\phi=0$, the constraint
equations on our initial data surface reduce to those of vacuum GR. 
For cases with a single black hole, we use either
harmonic~\cite{Cook:1997qc,Cook:2000vr} coordinates,
or Kerr-Schild \cite{PhysRevLett.11.237} coordinates
(which we discuss in more detail in Sec.~\ref{sec:single_bh}).
For constructing binary black hole initial data, we solve the Einstein constraints
using the conformal thin sandwich solver described in Ref.~\cite{East:2012zn}.

Given a particular choice of $H^a$, we need to ensure that the MGH 
condition, Eq.~\ref{eq:mh_condition},
is satisfied on the initial data surface.
Given initial data $\{g_{ab},\ \partial_0 g_{\alpha \beta}\}$
(and hence $\hat{g}_{ab}$ and $\tilde{g}_{ab}$,
see Eq.~\ref{eq:relation_gtilde_ghat_g})
that satisfy the constraints, we can always do this by solving
Eq.~\ref{eq:mh_condition} for $\partial_0 g_{0a}$.  In the language of the
$3+1$ decomposition, the choice of $H^c$ sets the initial time derivative of
the lapse function and shift vector.
\subsection{Diagnostic Quantities}

In order to characterize our results, we will make use
of several diagnostic quantities.
Considering first just the canonical coupling of the scalar field to gravity,
we can define a stress-energy
\begin{align}
    T_{ab}^{\rm SF}
   \equiv
   \frac{1}{8\pi}\left(
      \nabla_a\phi\nabla_b\phi-\frac{1}{2}g_{ab}\nabla_c \phi \nabla^c \phi
   \right) 
    \ ,
\end{align}
although note that when $\lambda\neq0$
this stress-energy is not generically conserved,
$\nabla^aT_{ab}^{\rm SF}\neq0$.
We can also define an effective stress energy tensor that is conserved,
simply by computing the Einstein tensor of the solution
\begin{align}
   T^{\rm Ein}_{ab}
   \equiv
    \frac{1}{8\pi}\left(R_{ab}-\frac{1}{2}R g_{ab}\right)
\end{align}
which would be equal to $T^{SF}_{ab}$ in the case that $\lambda=0$.
From $T^{\rm SF}_{ab}$ and $T^{\rm Ein}_{ab}$,
we can define effective energies and energy densities
\begin{align}
    E 
   \equiv
   \int t^a n^b T_{ab} \sqrt{\gamma}d^3x \equiv \int \rho_E \sqrt{\gamma}d^3x 
\end{align}
and angular momenta and associated densities:
\begin{align}
    J 
   \equiv
   \int \hat{\phi}^a n^b T_{ab} \sqrt{\gamma}d^3x \equiv \int \rho_J \sqrt{\gamma}d^3x 
\end{align}
where $t_a$ and $\hat{\phi}_a$ are, respectively,
the vectors pointing in the time and azimuthal directions,
which would be Killing vectors in the case that the spacetime is stationary and
axisymmetric.
We note that in axisymmetry, while $\rho_{J}^{\rm SF}$ will be identically zero,
$\rho_{J}^{\rm Ein}$ can actually be non-zero. 
Using these stress-energy tensors, we also define an energy flux through a surface as
\begin{align}
    \dot{E}
   \equiv
   \int -N t^a T_a^i dA_i \ .
\end{align}
We will mainly be interested in computing this quantity in the wavezone, 
at some surface at large radii. In that case, we expect $\dot{E}^{\rm Ein}$ to 
be the same as $\dot{E}^{\rm SF}$, due to the faster fall-off of the other curvature
terms in the wavezone. 

During the evolutions, we track any apparent horizons present at a given time,
and compute several diagnostic quantities with respect to them. From the area of
the apparent horizon, we can define an areal mass
$M_{\rm A}\equiv\sqrt{A/(16\pi)}$.
In a different context, this would be called the irreducible mass. However, the
spacetimes that we study here do violate the null convergence condition (which
states that  $R_{ab}k^ak^b\geq0$ for all null $k^a$), and thus there will be
cases where $M_{\rm A}$ decreases. We can also associate an angular momentum
to the apparent horizon\footnote{We recall that this quantity is conserved
if $\hat{\phi}_i$ is tangent to a Killing vector field, regardless of
whether the Einstein equations hold, or if the spacetime obeys any
energy conditions \cite{poisson2004relativist}.} 
\begin{align}
   \label{eq:j_ah}
   J_{\rm AH}
   \equiv
   \frac{1}{8\pi} \int \hat{\phi}_i K^{ij} dA_j \ , 
\end{align}
and using the Christodoulou formula, a mass
\begin{align}
   \label{eq:m_ah}
    M_{\rm AH} 
   \equiv
   \left( M_A^2+\frac{J_{\rm AH}^2}{4M_A^2}\right)^{1/2}  \ .
\end{align}
As an indication of the scalar hair formation about the black hole,
we also keep track of the area
averaged value of the scalar field on
the apparent horizon $\langle \phi \rangle_{\rm AH}$.
In order to compute the gravitational radiation, we extract
the Newman-Penrose scalar $\psi_4$.

\section{Results}
\label{sec:results}
\subsection{Hyperbolicity tests with weak field data}
\label{sec:hyperbolicity_tests}

As a first test, we consider a weak field configuration, and
provide numerical evidence that
the equations of motion for shift-symmetric ESGB gravity are strongly
hyperbolic in the MGH formulation, with the gauge choices we have made.
A necessary condition for ESGB gravity to have a well-posed initial value problem
is for the equations of motion to have a strongly hyperbolic formulation (e.g.
\cite{Sarbach:2012pr}). Papallo and Reall~\cite{Papallo:2017qvl,Papallo:2017ddx} have
shown that in the generalized harmonic formulation, the EOMs for ESGB gravity
are not strongly hyperbolic around generic weak field solutions, instead they
are only weakly hyperbolic.  Later, Kovacs and Reall showed that the equations
of motion for ESGB gravity are strongly hyperbolic in the MGH formulation~\cite{Kovacs:2020ywu},
for weak coupling backgrounds where all the characteristic length scales
(associated with the spacetime curvature and scalar gradients)
satisfy $L \gg \sqrt{\lambda}$.

In general, one expects that a set of nonlinear weakly (but not strongly)
hyperbolic equations of motion will have modes that exhibit frequency dependent
growth, where the growth rate increases as a polynomial in the frequency.
Given this, we expect that simulations of ESGB gravity in a generalized harmonic
formulation should generally not converge with higher resolution, since the
higher resolution will resolve smaller scales, and thus allow faster growing
fluctuations.  This being said, with sufficiently smooth initial data (in
particular without AMR, moving excision surfaces, etc.,
which tend to introduce high frequency numerical error), at a given
\emph{fixed} resolution, it may be difficult to observe small scale growth over
a finite simulation run time, and one must use non-smooth initial data in order to make
this problem apparent (see e.g. \cite{Giannakopoulos:2020dih}
and references therein). This is the approach we take here.

Usually, hyperbolicity or robustness tests in GR are performed around Minkowski
space or other trivial, scale free background solutions.  However, the analysis
in Ref.~\cite{Papallo:2017ddx} indicates that a ``generic" background solution
that violates strong hyperbolicity for the generalized harmonic formulation
requires non-vanishing derivatives of the scalar field. Therefore, we must
resolve a hierarchy of scales given by $L \gg \sqrt{\lambda} \gg \omega^{-1}$,
where $L$ is the characteristic length scale of the background curvature,
and $\omega$ is the frequency of the modes which may violate strong
hyperbolicity\footnote{We recall that we need $L \gg \sqrt{\lambda}$, as we are
considering hyperbolicity in the weak field regime--in the strong
field regime it is likely the theory is not even weakly hyperbolic
\cite{Ripley:2019hxt,Ripley:2019irj}.}. 
In order to make reaching these high resolutions tractable, we
impose a translational symmetry in the $z$ direction, and restrict to a two
dimensional, periodic domain of length $L$. The initial data we consider is as
follows. For the scalar field we set 
\begin{align}
    \phi(t=0)=\bar{\phi}\sin(2\pi x/L)\sin(2\pi y / L) , \ \partial_0 \phi(t=0)=0
\end{align}
where here we take the amplitude to be $\bar{\phi}=0.01$.
We set the metric to be initially Minkowski, but add a small white noise 
perturbation to the initial metric time derivative
\begin{align}
     g_{ab}(t=0) = \eta_{ab}, \ \partial_0 g_{ab}(t=0) = f \mathcal{N}/L  \ .
\end{align}
Here $\mathcal{N}$ gives a random number between -1 and 1 at every spatial
point, and $f$ is a constant controlling the amplitude. Even if $f$ were zero,
this solution does not satisfy the constraints, though the constraint violation
will in some sense be small since $\bar{\phi} \ll 1$, and our goal here is
merely to study the hyperbolicity of the free evolution equations\footnote{Moreover
we note that,
the hyperbolicity analysis of Horndeski gravity theories in generalized
harmonic formulations in Ref.~\cite{Papallo:2017qvl,Papallo:2017ddx} indicate that the weakly
hyperbolic modes were constraint violating ones; thus we expect that the differences
between simulations in the MGH and generalized harmonic formulations
will be most apparent when we begin with initial data which slightly
violate the constraint equations.}. 
We only perturb $\partial_0 g_{ab}$ to avoid any issues with the evolution
equations being second order in $g_{ab}$ (as constructing a first order
version would require introducing a new evolution variable constrained to be
equal to $\partial_\alpha g_{ab}$).
We set $\lambda/L^2=0.025$, so that we are in the weak coupling regime.

We consider a sequence of numerical evolutions where we simultaneously increase
the resolution for the numerical grid, while decreasing $f$. In particular, we
consider grid spacings $dx=dy=h_i=L/2^{7+i}$ with $i=0$, 1, \ldots, 5, and
scale the white noise amplitude as $f_i=f_0/2^{4i}$ so that it scales in the
same manner as the truncation error with our fourth-order scheme. We choose
$f_0$ so that the white noise perturbation is small, but dominates over the
truncation error ($f_0\approx 5\times 10^{-6}$). To monitor the subsequent
behavior of the high frequency perturbation, we compute the following difference
between subsequent resolutions 
\begin{align} ||\partial_0 g_{ab}^{h} || =
\sum_{ab} \left[ \sum_{x,y} \left(\partial_0 g_{ab}^{h} -\partial_0
g_{ab}^{h/2}\right)^2 \right]^{1/2} 
    \label{eq:hyper_norm}
\end{align} 
where $\partial_0 g_{ab}^{h}$ is the numerical solution computed with grid
spacing $h$, the outer sum is a sum over the 10 unique metric time derivatives,
and the inner sum is a sum over the points in the $x$ and $y$ directions
(restricted to a coarse grid of points shared by all resolutions).

\begin{figure*}
\begin{center}
    \includegraphics[width=0.66\columnwidth,draft=false]{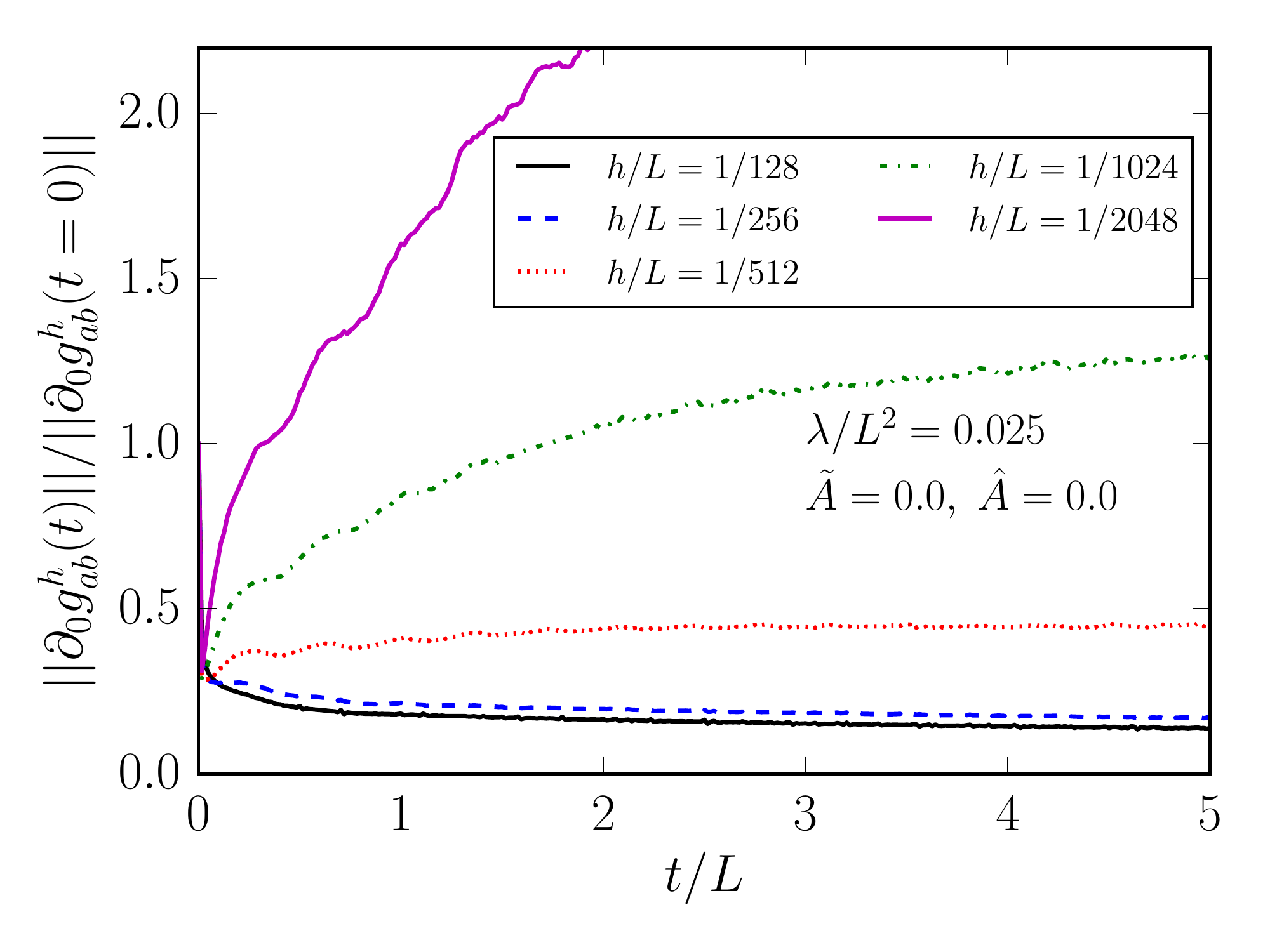}
    \includegraphics[width=0.66\columnwidth,draft=false]{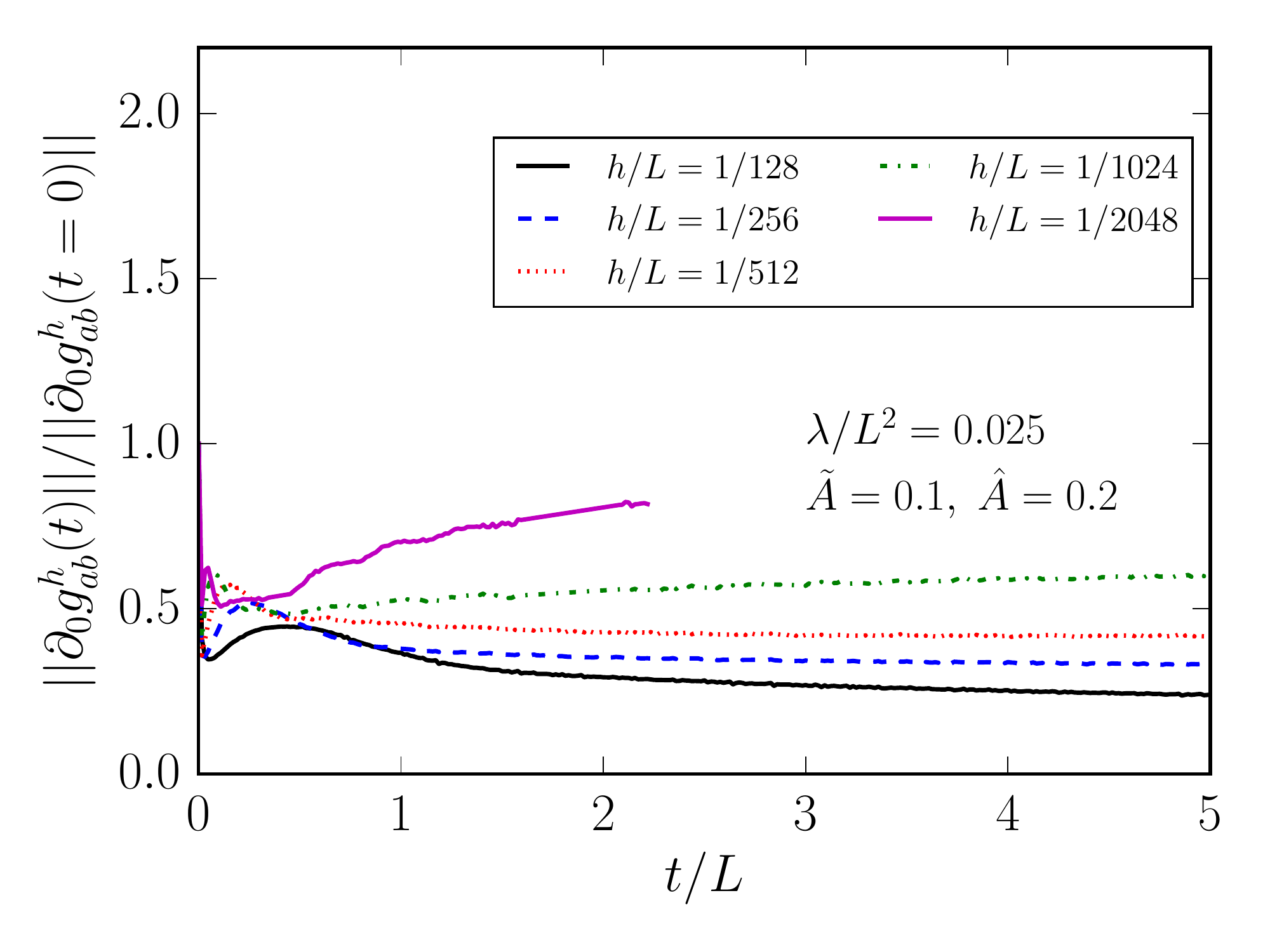}
    \includegraphics[width=0.66\columnwidth,draft=false]{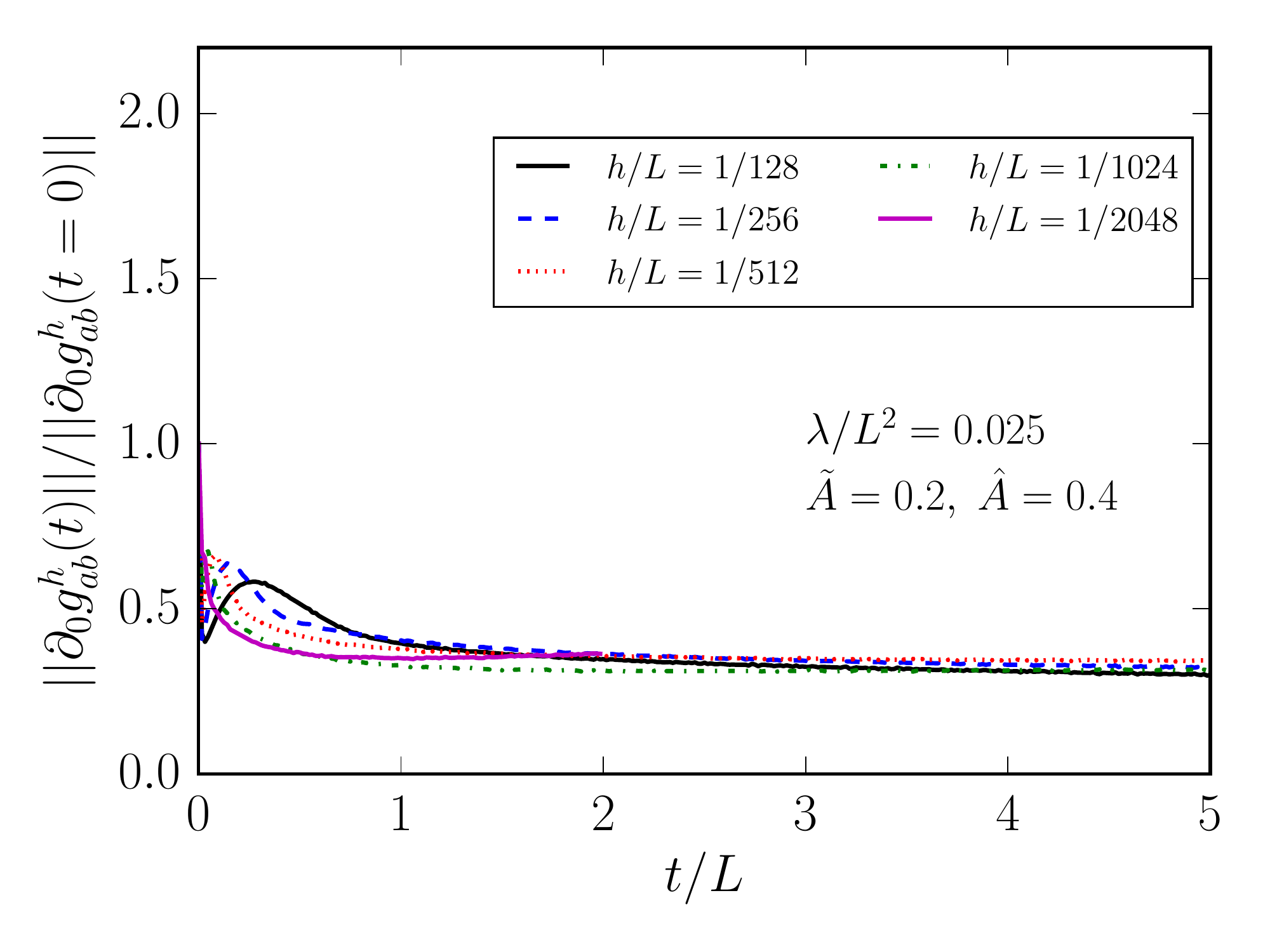}
\end{center}
\caption{
    We show the difference in the evolution variables $\partial_0 g_{ab}$
    between subsequent resolutions, computed using Eq.~\ref{eq:hyper_norm}, and
    normalized by the initial value of the white noise perturbation, as a function
    of time, for a periodic scalar field configuration.  The three cases correspond
    to different choices of the auxiliary metrics in MGH with (left to right)
    $(\tilde{A},\hat{A})=(0,0)$, $(0.1,0.2)$, and $(0.2,0.4)$.  In the first case,
    corresponding to harmonic gauge, frequency dependent growth can be clearly seen
    as higher frequencies perturbations are sourced for the higher resolutions.
    With auxiliary metrics that differ sufficiently from the physical one, as in
    the right-most panel, this problem no longer appears.  Note that the
    $h/L=1/4096$ resolution run (needed to compute the error of the $h/L=1/2048$ run)
    was continued for a shorter time compared to the lower resolutions due to
    computational expense. 
\label{fig:hyper_test}
}
\end{figure*}

We restrict to a gauge with $H^a=0$, but consider three different choices for
the auxiliary metrics corresponding to Eq.~\ref{eq:relation_gtilde_ghat_g} with
$(\tilde{A},\hat{A})=(0,0)$ (harmonic gauge), $(\tilde{A},\hat{A})=(0.1,0.2)$,
and $(\tilde{A},\hat{A})=(0.2,0.4)$.  We show the results for these cases in
Fig.~\ref{fig:hyper_test}. When the auxiliary metrics are set equal to the
physical metric, we do indeed find a perturbation that grows faster and faster
as the grid spacing, and hence the minimum wavelength of the perturbation, is
decreased. Changing the lightcones for the auxiliary metrics by using nonzero
$(\tilde{A},\hat{A})$ improves this, and for $(\tilde{A},\hat{A})=(0.2,0.4)$
there is no evidence of frequency dependent growth.

There is of course no requirement that $\tilde{A}$ and $\hat{A}$ be of some fixed ratio, 
and in general there is a large degree of freedom in choosing the auxiliary metrics
which we do not systematically explore here. For this study, we concentrate merely
on finding a choice of parameters that works, and
for most of the remaining applications, we will use
$(\tilde{A},\hat{A})=(0.2,0.4)$, and rely on convergence tests to estimate the
accuracy of our results, and as a check for
contamination of the solution due to ill-posedness.

\subsection{Single black hole initial data}
\label{sec:single_bh}

   We next present simulations where out initial data
is a single black hole, restricting to axisymmetry.
As discussed in Sec.~\ref{sec:initial_data},
we begin our evolution in Kerr-Schild coordinates \cite{PhysRevLett.11.237}.
Our main conclusion in this section is that in full shift-symmetric ESGB gravity
(for small enough coupling parameters $\lambda$),
Kerr initial data leads to stable, rotating, scalar hairy black hole solutions.
We note that Kerr solutions for ESGB gravity
were evolved using an order-reduction approach in 
Ref.~\cite{Okounkova:2019zep},
and stationary solutions to the full theory describing
spinning black holes with scalar hair were constructed
in Refs.~\cite{Delgado:2020rev,Sullivan:2020zpf}.

\begin{figure*}
\begin{center}
    \includegraphics[width=1.0\columnwidth,draft=false]{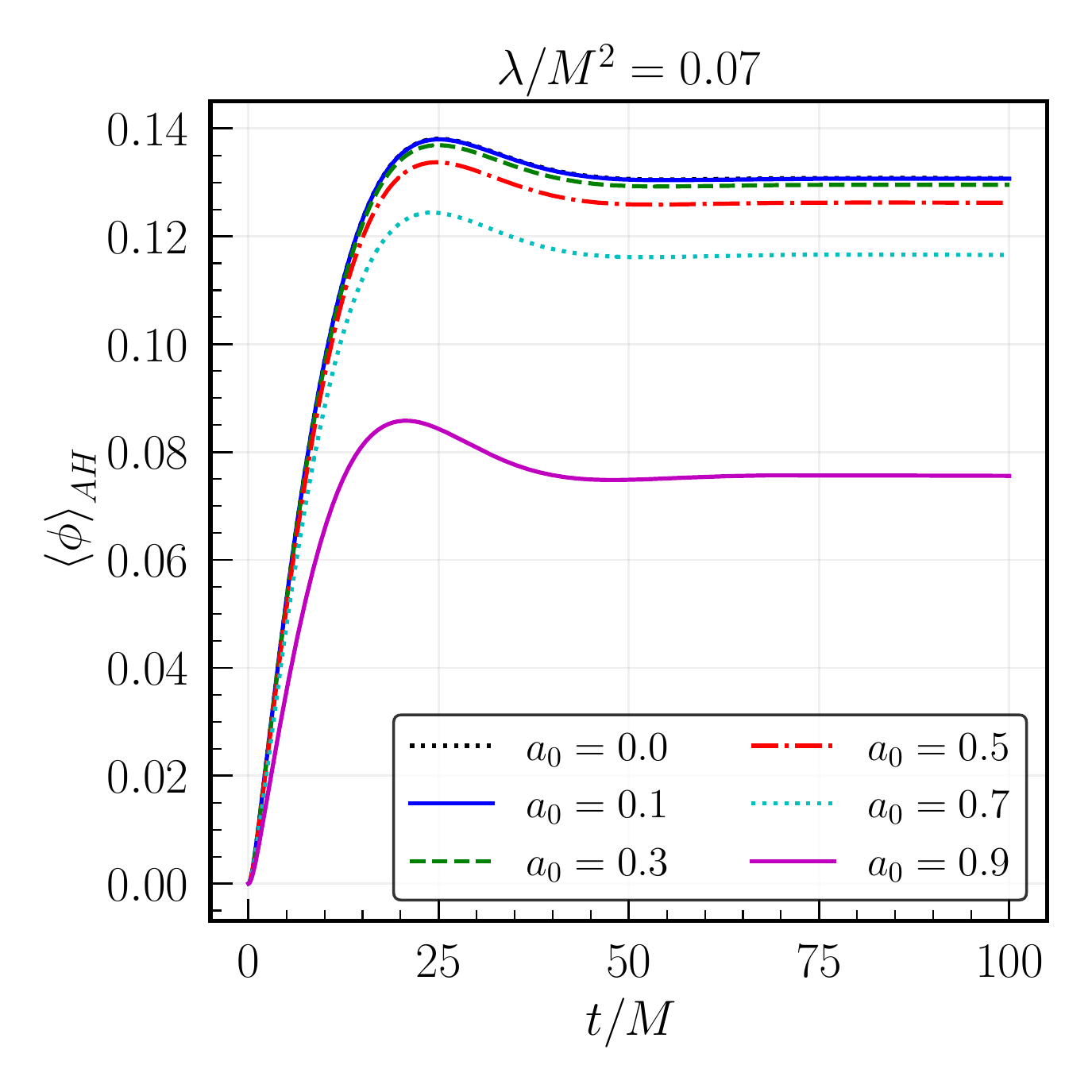}
    \includegraphics[width=1.0\columnwidth,draft=false]{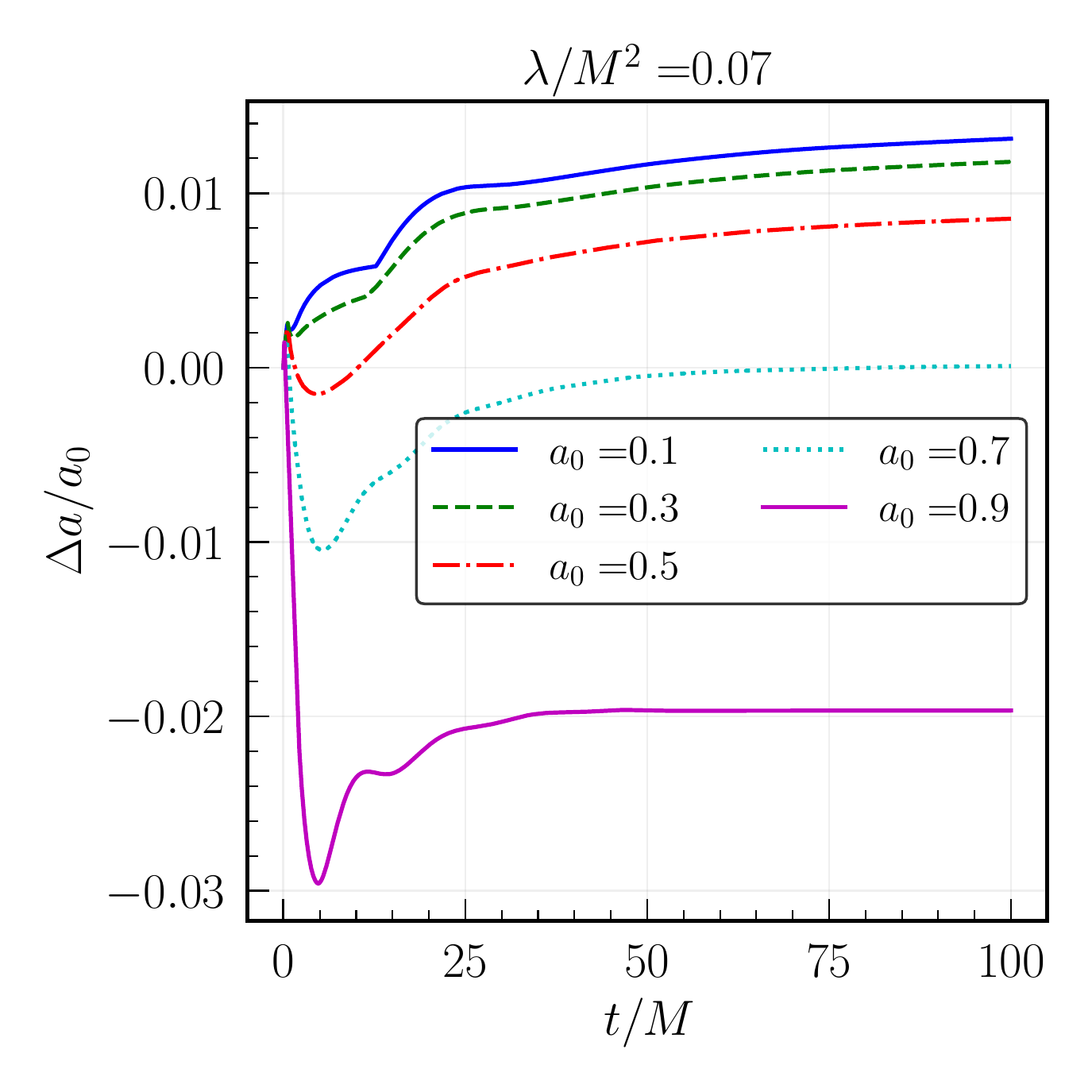}
    \includegraphics[width=1.0\columnwidth,draft=false]{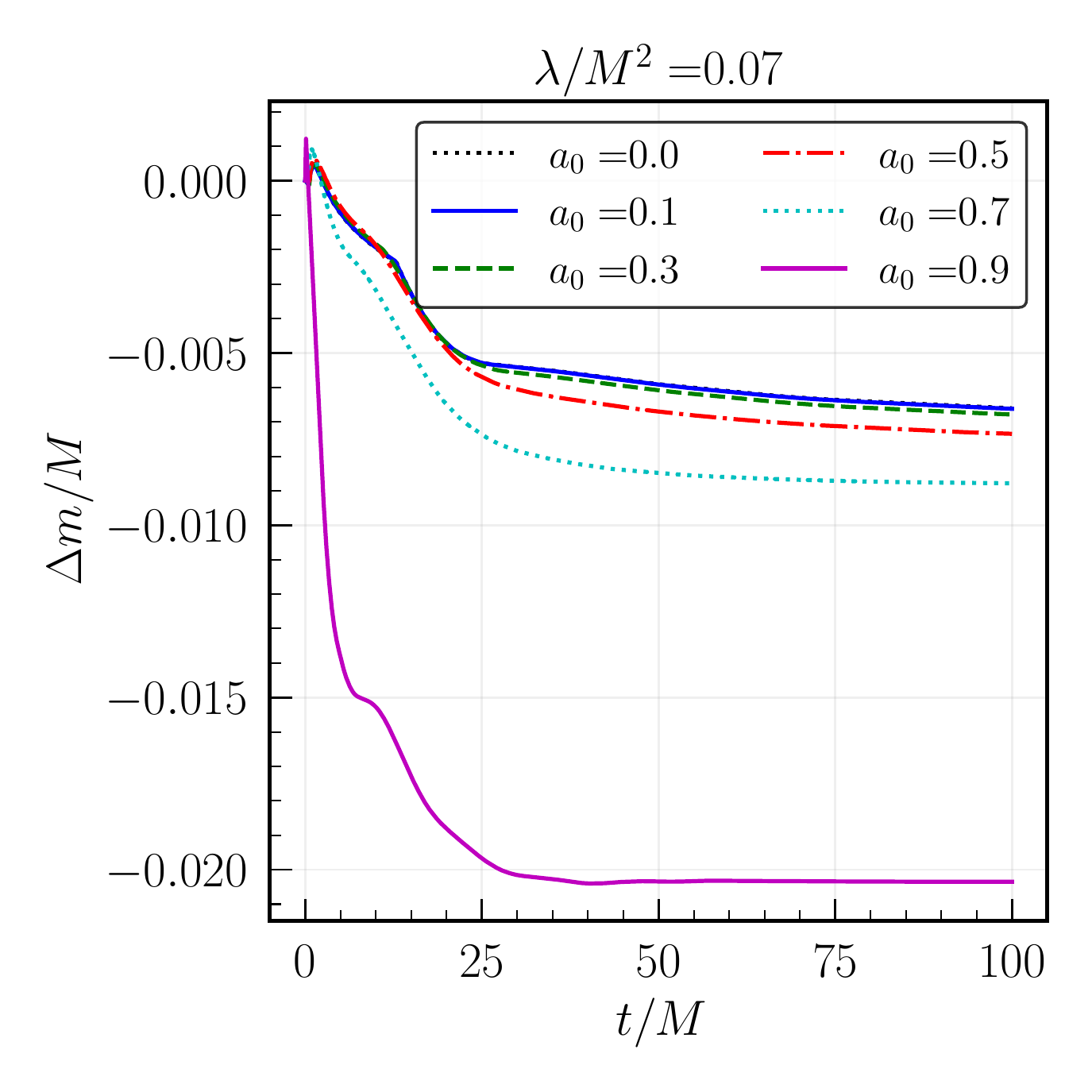}
\end{center}
\caption{
   Left to right: The horizon-averaged scalar field $\left<\phi\right>_{AH}$,
value of dimensionless black hole spin (normalized by the initial black hole
spin $a_0$), and value of black hole mass $m$ (normalized by the ADM mass $M$),
for a one-parameter family of Kerr initial data.  We show the
scalar cloud profile on the black hole horizon for
$\lambda/M^2=0.07$ (runs with $\lambda/M^2=-0.07$ give qualitatively similar
results, except $\left<\phi\right>_{AH}$ has the opposite sign).  Though the
black hole always loses angular momentum as a result of
scalar hair growth, from the
center panel we see that for small enough black hole spins ($a\lesssim0.7$),
the dimensionless spin decreases, while for larger spins it increases (for the
case $a_0=0$, the black hole spin does not change, so we omit it from this
figure).  The growth of the scalar cloud always coincides with
a decrease in the mass of the black hole,
which we interpret as the scalar field extracting energy from the black
hole. 
}
\label{fig:black_holes_spin_lte_09}
\end{figure*}

In Fig.~\ref{fig:black_holes_spin_lte_09}, we plot the average scalar field
value over the black hole apparent horizon $\left<\phi\right>_{AH}$,
along with the change in the black hole mass and spin
and the change in black hole mass and spin as measured
on the horizon (see~Eqs.~\eqref{eq:j_ah} and ~\eqref{eq:m_ah}), as a
function of evolution time $t$, for different initial (dimensionless) black
hole spin parameters $a_0$, given a fixed value of $\lambda/M^2=0.07$ (where
$M$ is the Arnowitt-Deser-Misner (ADM) mass). 
For a given black hole spin, Ref.~\cite{Delgado:2020rev}
found that there is a maximum $\lambda/M^2$ above which they could no longer
construct regular solutions to the theory, and this maximum decreased
with increasing spin.  For a spin parameter of $a=0.9$,
the coupling limit was found to be $\lambda/M^2\approx 0.13$,
while for $a=0.99$,
it was found to be roughly a factor of two smaller~\cite{Delgado:2020rev}.
We find it difficult to consider couplings near this limit for a given spin,
which we believe is in part due to the fact that the scalar hair
initially exceeds its
stationary value during the growth of the scalar cloud before settling down to a
lower value (e.g. the top left panel of Fig.~\ref{fig:black_holes_spin_lte_09};
see also \cite{Ripley:2019aqj}), so that any problems that could occur from,
e.g. loss of hyperbolicity come closer to the horizon than one would anticipate
from studies of stationary solutions.
But we have not explored extensively whether better choices
of gauge and auxiliary metrics could also improve this.  

We find that the average asymptotic scalar field value at the horizon decreases
as a function of initial dimensionless Kerr spin parameter.
This is consistent with the fact that the average value of the
Gauss-Bonnet scalar is a decreasing function of black hole spin;
in fact, at a critical value of $a\gtrsim0.766$,
the average value of the Gauss-Bonnet scalar is negative
on the black hole horizon.
We find that the average value of the scalar field only becomes negative
for larger spins ($a\gtrsim0.95$), which is most likely due to the fact
that stationary scalar field configurations must balance gradients with the
varying Gauss-Bonnet source term on the horizon.
We find that as we increase the black hole 
spin, the scalar field becomes negative on the spin axis (where the
Gauss-Bonnet scalar is negative), but remains
positive on the equator of the black hole (where the Gauss-Bonnet scalar is positive). 
The formation of scalar hair decreases both the mass and angular momentum of 
(as measured on the horizon) of the initial spinning black hole.
However, from Fig.~\ref{fig:black_holes_spin_lte_09},
we can see that for initial black hole spins that are roughly less than
$a_0\lesssim0.7$,
black hole scalar hair formation increases the dimensionless spin, while
for greater initial black hole spins, it decreases the dimensionless spin 
somewhat.
For initial black hole spin $a=0.9$ and $\lambda/M^2=0.07$, we see
that the change in the dimensionless black hole spin is approximately
$\sim -2\%$ while the change in the black hole mass is approximately $\sim-2\%$.

\begin{figure*}
\begin{center}
    \includegraphics[width=0.9\columnwidth,draft=false]{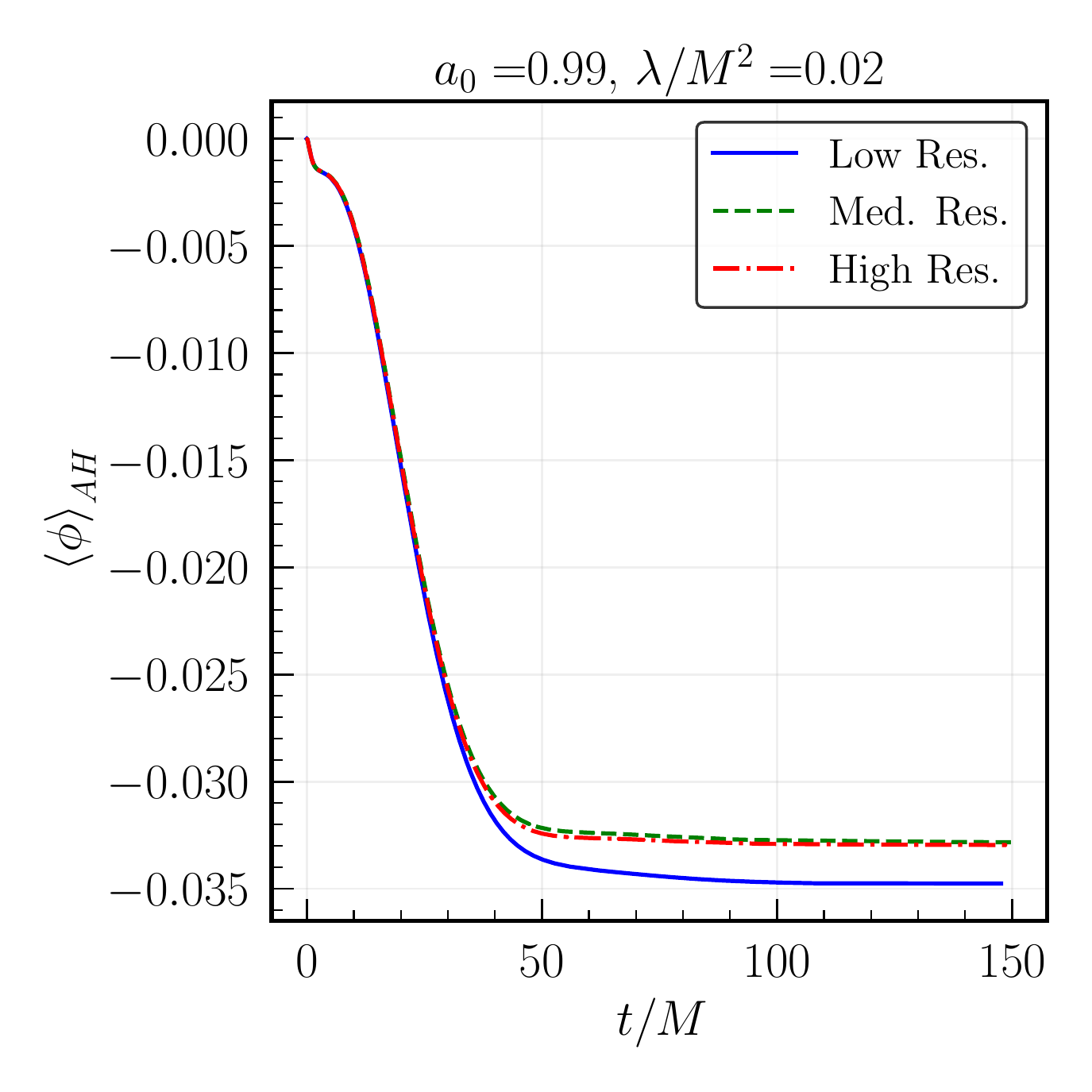}
    \includegraphics[width=0.9\columnwidth,draft=false]{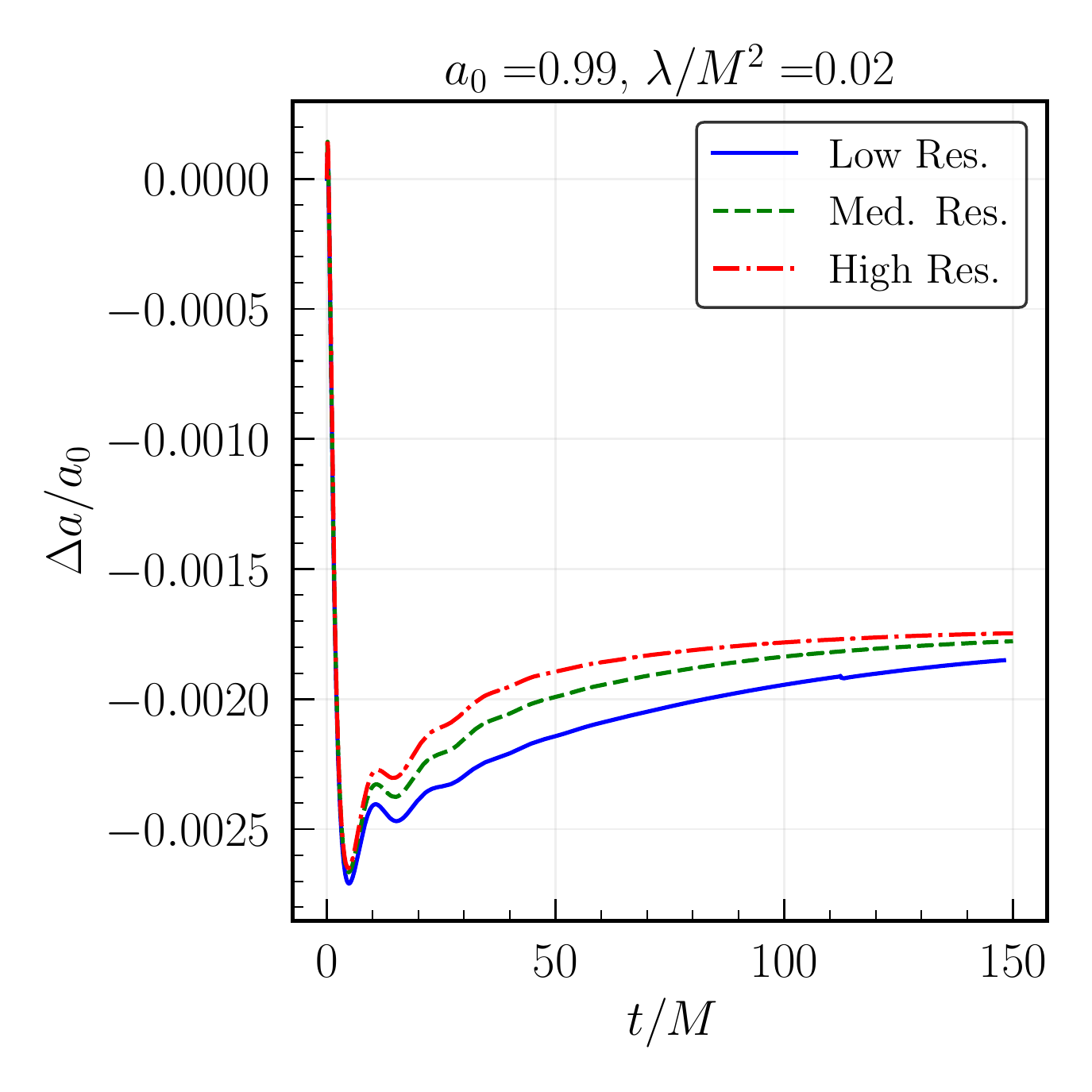}
    \includegraphics[width=0.9\columnwidth,draft=false]{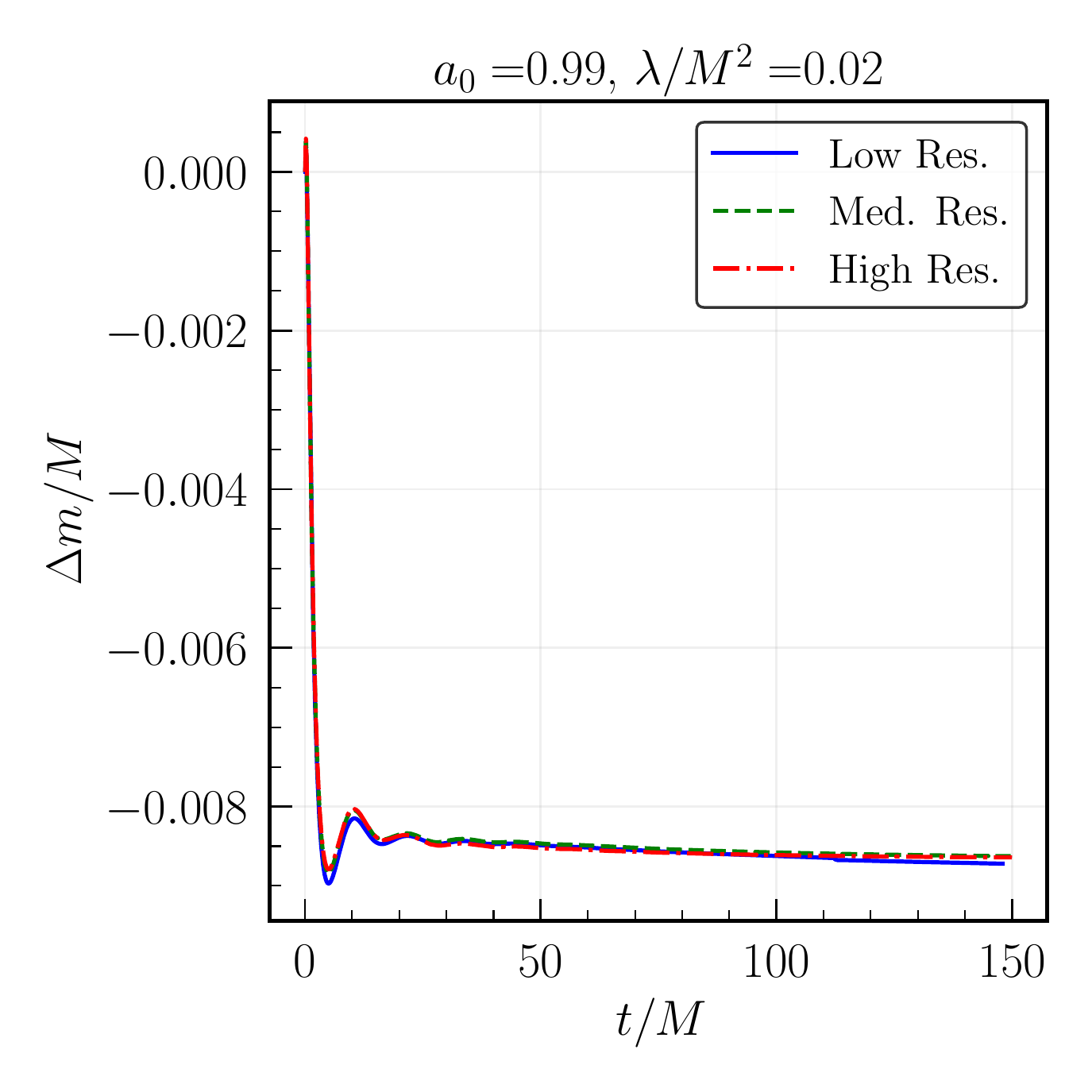}
    \includegraphics[width=0.9\columnwidth,draft=false]{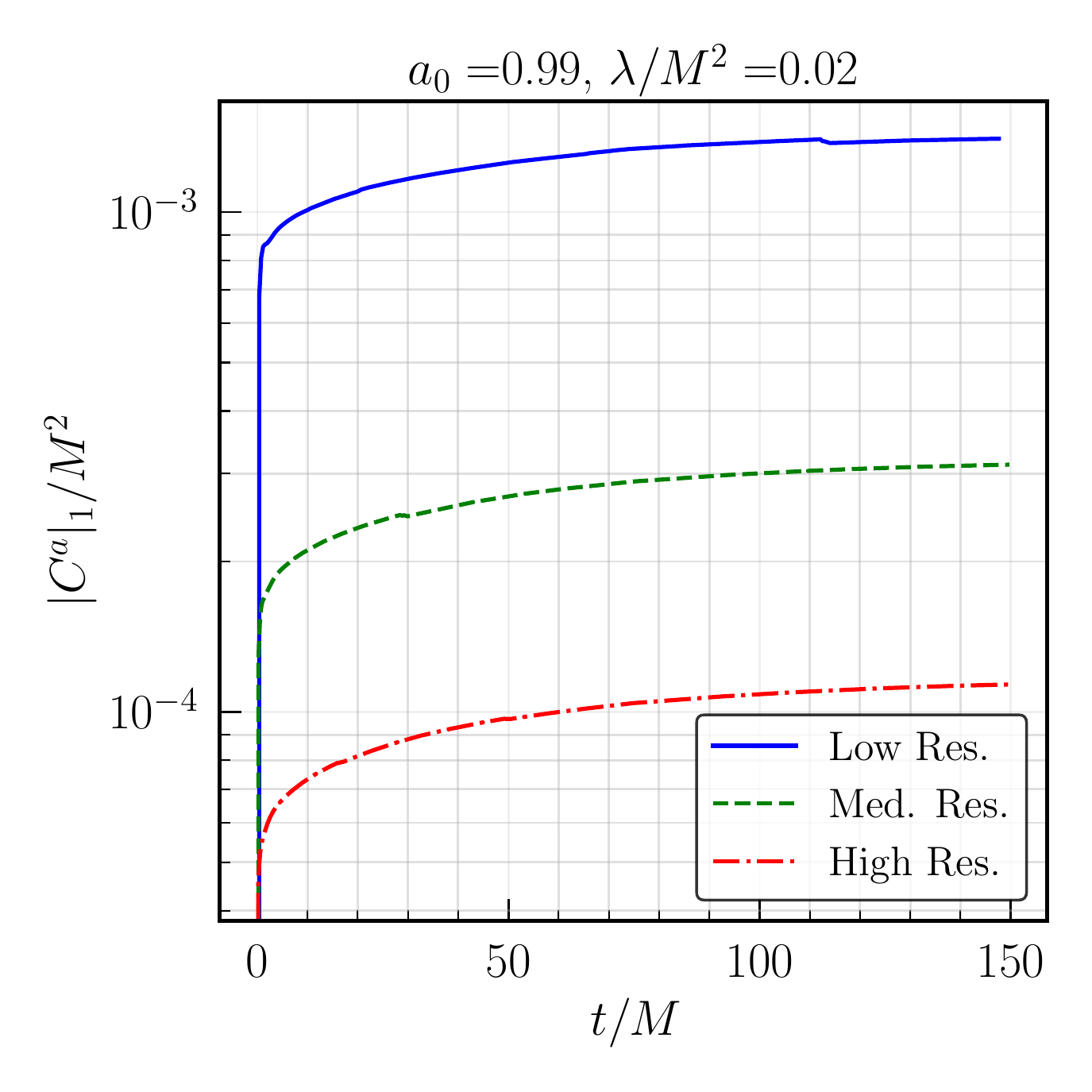}
\end{center}
\caption{
  Convergence study of the scalar hair growth about a black hole
with $a_0=0.99$ and $\lambda/M^2=0.02$. We show the horizon-averaged scalar field
$\left<\phi\right>_{AH}$ (top right), change in dimensionless black hole spin $a$ (top left),
relative change in the black hole mass (bottom right), and 
constraint violation $|C^a|$ (normalized
by the initial black hole mass $M$; bottom left).
We find that the constraint violation converges at third order,
as expected. 
The medium and high
resolutions have $1.5$ and $2\times$ the linear resolution of the low
resolution simulation.
For an image of the scalar field density around
the black hole, see Fig.~\ref{fig:scalar_field_density}.
}
\label{fig:black_holes_spin_099}
\end{figure*}

\begin{figure}
\begin{center}
    \includegraphics[width=0.9\columnwidth,draft=false]{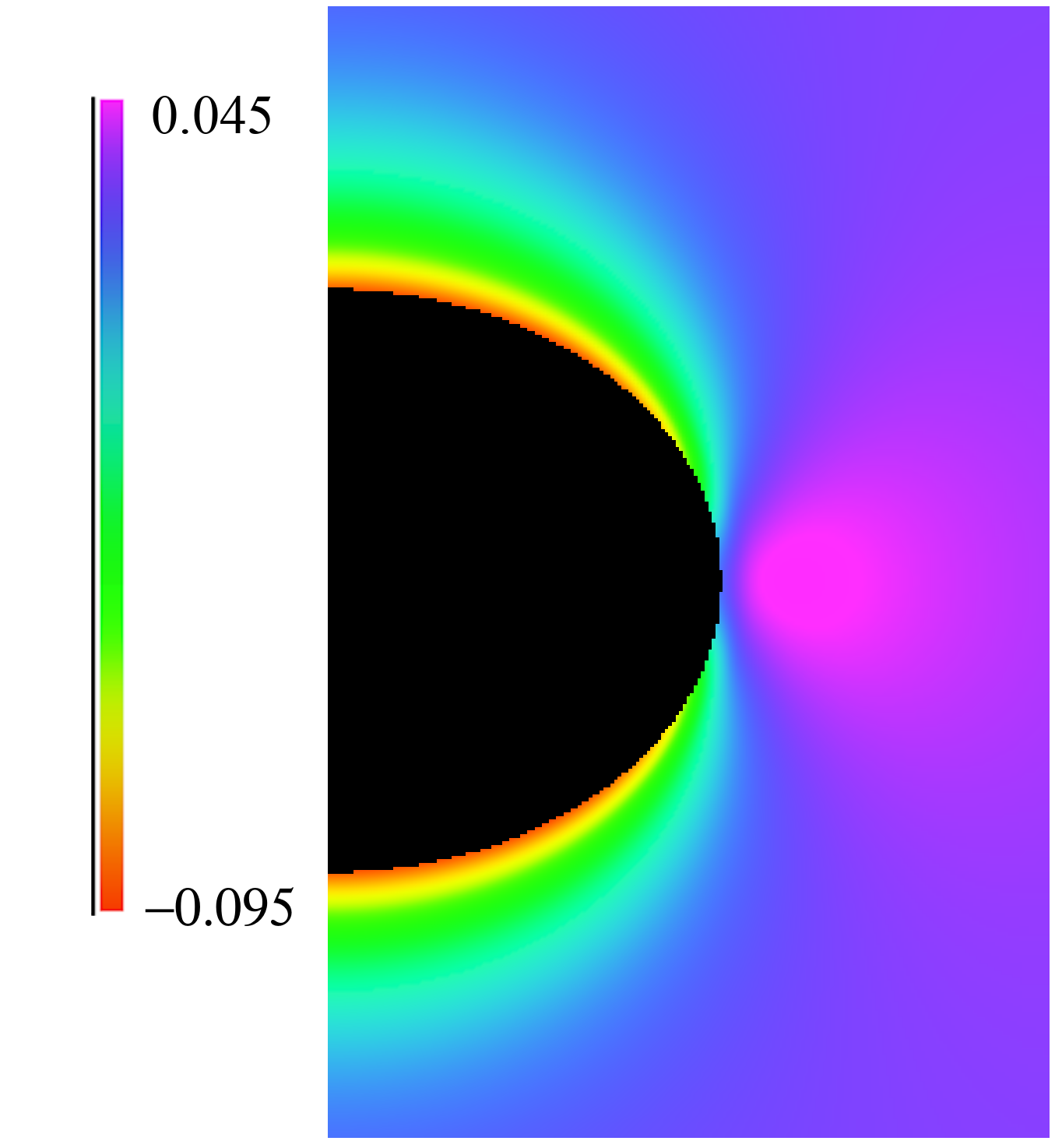}
\end{center}
\caption{
The scalar field value around a black hole of initial dimensionless spin
$a_0=0.99$ and dimensionless coupling $\lambda/M^2=0.02$, taken
after $150M$ of evolution.  For the evolution of the black hole parameters, see
Fig.~\ref{fig:black_holes_spin_099}.  The spacetime is axisymmetric, and we
show a slice at fixed azimuthal angle, with the bottom of the figure
corresponding to the axis of symmetry.  The excised region
within the apparent horizon, which is roughly $90\%$ of the radius of the black
hole, is shown in black.  The colorbar indicates the scalar field value,
which in this case varies between $\phi\in[-0.095,0.045]$.  We see that on the
equator of the spinning black hole (middle of the figure), the scalar field is
positive, while at the poles, the scalar field is negative.  This is to be
contrasted with black holes with zero spin and $\lambda>0$, where $\phi$ is
everywhere positive.
}
\label{fig:scalar_field_density}
\end{figure}

In Fig.~\ref{fig:black_holes_spin_099}, we show a convergence
study of the average scalar hair profile, along with the
change of the black hole angular momentum and
mass (see~Eqs.~\eqref{eq:j_ah} and ~\eqref{eq:m_ah})
as measured on the horizon of a black hole with initial dimensionless
spin parameter of $a_0=0.99$.
We see that the change in the black hole horizon angular momentum
is $\sim0.2\%$, while the change in the black hole mass is $\sim0.8\%$.
In this study, the integrated constraint violation $C^a$ converges at third order,
(consistent with the time interpolation used by the AMR algorithm).
Here, and in subsequent sections, we show $|C^a|$ integrated over the 
coordinate radius $r\leq 100\ M$ region of the domain.
We found that as we considered larger black hole spins, we could only obtain
stable, convergent evolution with small ESGB couplings, and
had to place our excision radius closer to the black hole horizon.
Figure \ref{fig:scalar_field_density} shows a snapshot
of the scalar field around a black hole
with initial spin $a=0.99$, taken after $150M$ of evolution.
The scalar field is positive around the equator
of the black hole, while it is negative around the spin axis.

One reason to expect that only smaller Gauss-Bonnet couplings can be
used to evolve higher spin black holes is because
the Kerr ring curvature singularity moves closer
to the black hole horizon for larger spins, and there is
numerical evidence that the equations of motion for ESGB gravity
are hyperbolic only for regions of relatively small curvature
(given a fixed Gauss-Bonnet coupling)
\cite{Ripley:2019hxt,Ripley:2019irj,Ripley:2019aqj}.

\subsection{Head-on binary black hole mergers}
\label{sec:axisym_bhbh}
We next study binary black hole mergers in ESGB. We begin by restricting
to the axisymmetric case of a head-on collision, which allows us to quickly
cover a number of different parameters, including different values of $\lambda$,
as well as different black hole spins and mass ratios. Since the corrections
in ESGB are sensitive to the smallest length scale, we will label the cases
we consider in terms of the quantity $\lambda/m^2$, where $m$ is the mass of the smallest black hole
in the initial data.

Our main result in this section is that we find that the ESGB theory in
general, and the MGH formulation in particular, remains hyperbolic, even in the
highly dynamical setting of a (head-on) binary black hole merger, for
comparable values of $\lambda$ to where the spherically symmetric problem
remains well posed. For reference, in Ref.~\cite{Ripley:2019aqj}, the maximum
value where the scalar hair grew about a
Schwarzschild black hole that could be evolved without
the loss of hyperbolicity was $\lambda/m^2\approx 0.19$, and based on
extrapolation, it was estimated that hyperbolicity would be lost outside the
black horizon for $\lambda/m^2\gtrsim 0.23$.

\begin{figure}
\begin{center}
   \includegraphics[width=\columnwidth,draft=false]{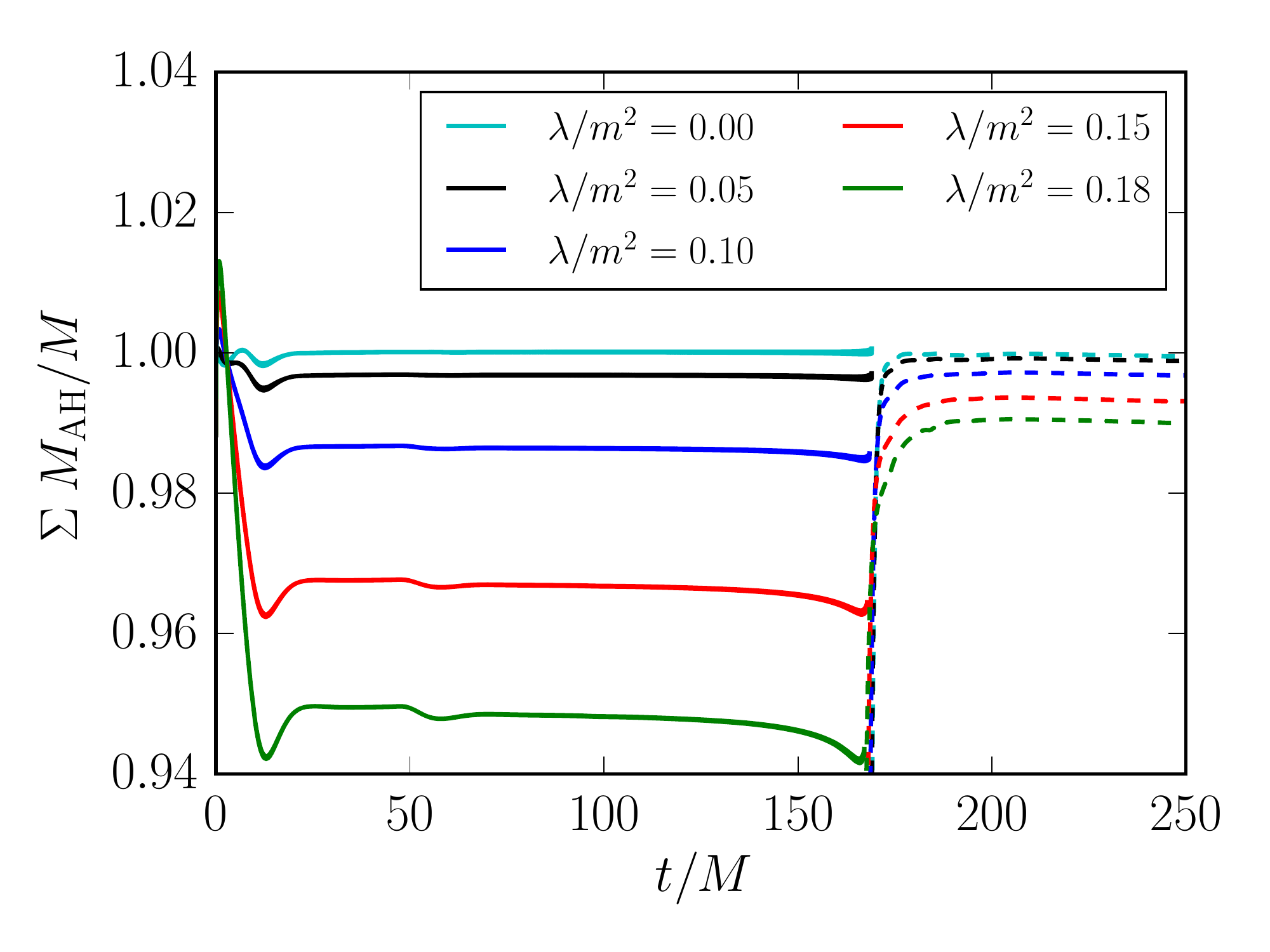}
   \includegraphics[width=\columnwidth,draft=false]{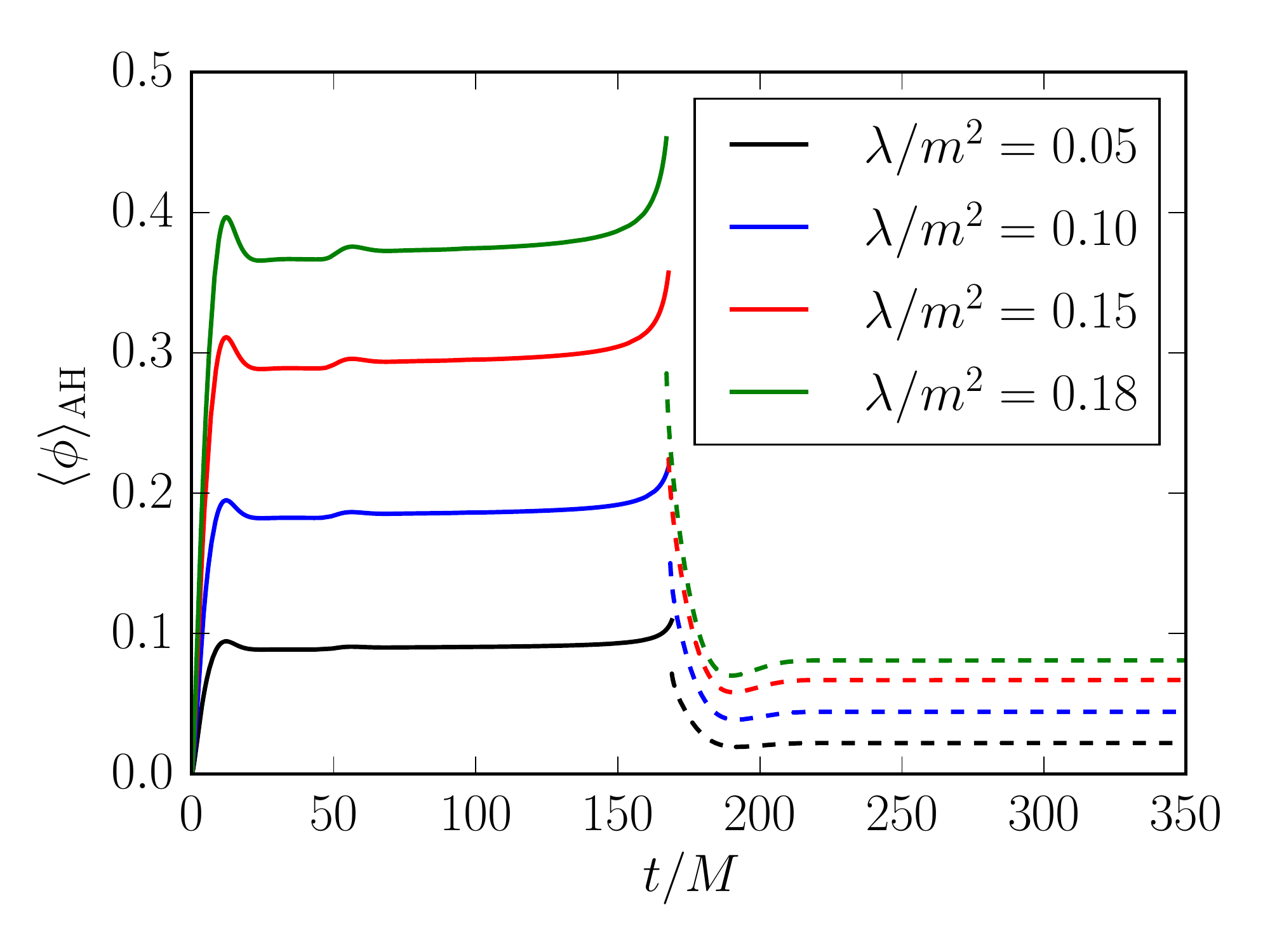}
\end{center}
\caption{
    Top: The sum of the masses of the merging black holes (solid lines)
    or mass of the final black hole (dashed lines)
    as a function of time for head-on mergers of equal mass,
    non-spinning black holes and different values of $\lambda$. 
    Bottom: The area-averaged value of $\phi$ on the apparent horizon as a function of time.
\label{fig:bhbh_mbh_comp}
}
\end{figure}

As discussed in Sec.~\ref{sec:initial_data},
we start with initial data where $\phi$ and $\partial_0
\phi$ are identically zero. Hence, initially the individual black holes will develop
scalar hair as they fall towards each other and finally merge.  We choose the
initial separation of the black holes to be $d=50M$ (where $M$ it the ADM mass of the
spacetime), and set their initial velocities to the value corresponding to the
binary being marginally bound.  We show a number of cases with an equal-mass,
non-spinning binary black hole, and different values of the coupling ranging from $\lambda/m^2=0$ to
0.18 in Fig.~\ref{fig:bhbh_mbh_comp}.
Initially scalar hair grows about the black hole, which loses mass
as the cloud grows.  For larger couplings, there is a small increase
in the magnitude of $\phi$ on the horizons as the black holes approach each other
(bottom panel of Fig.~\ref{fig:bhbh_mbh_comp}), and corresponding decrease in
black hole mass (top panel). However, when the black holes merge, forming a larger black hole, the
Gauss-Bonnet curvature outside the common horizon becomes smaller, and the
scalar cloud shrinks. As elaborated on below, most of the energy lost by
the smaller black hole goes back into the remnant black hole,
as opposed to escaping as radiation.

\begin{figure}
\begin{center}
    \includegraphics[width=\columnwidth,draft=false]{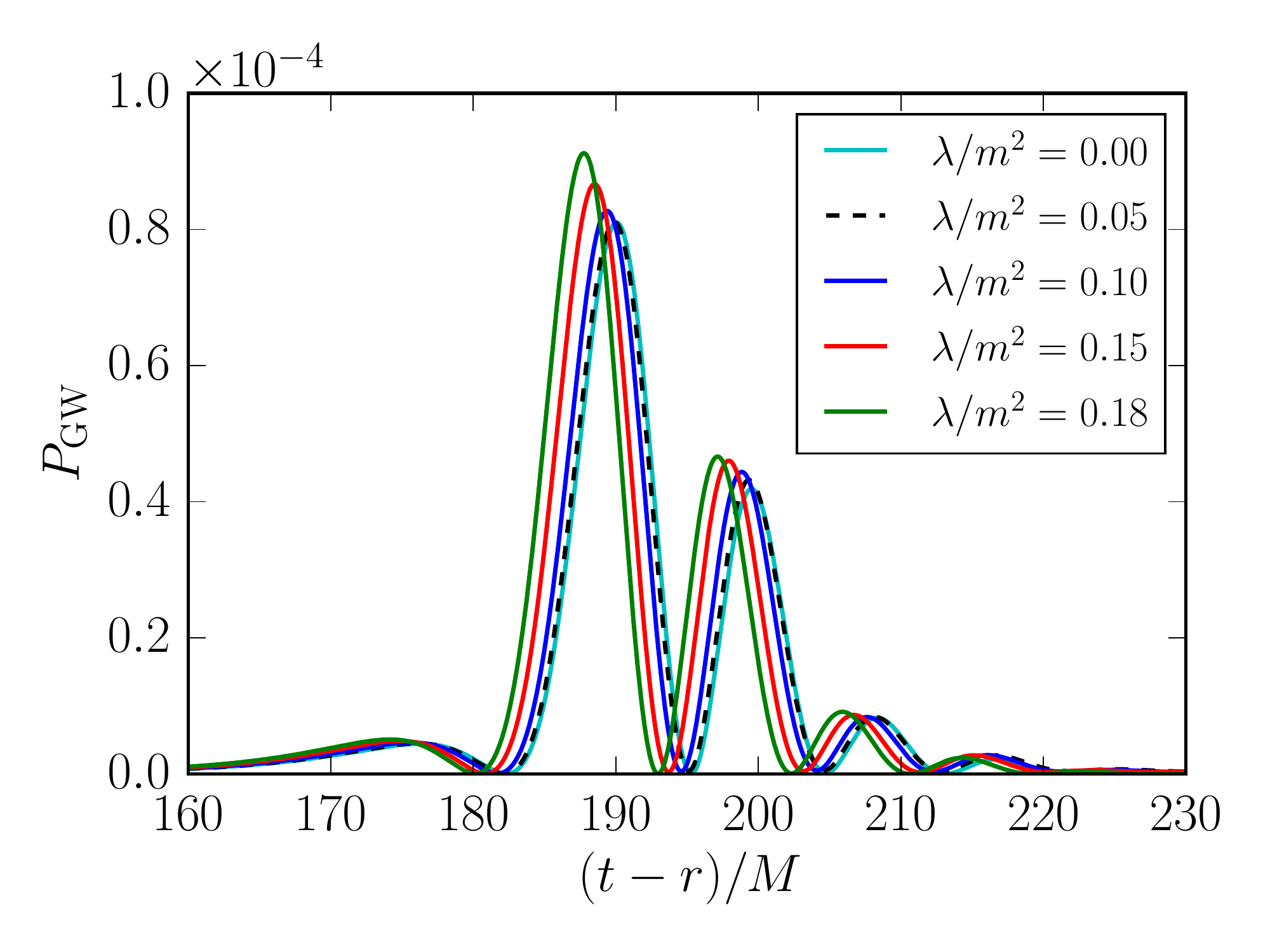}
    \includegraphics[width=\columnwidth,draft=false]{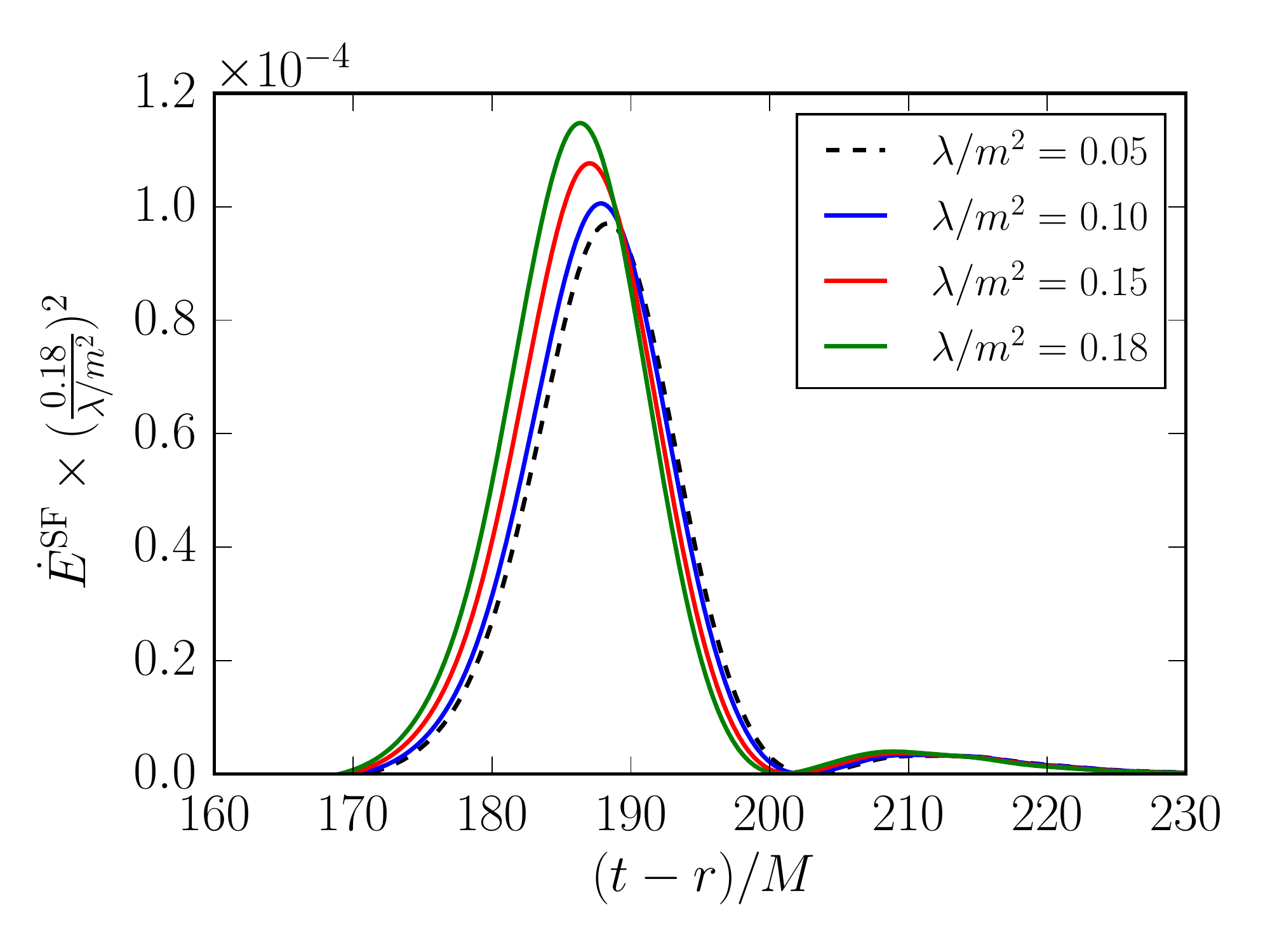}
\end{center}
\caption{
    Top: The gravitational wave luminosity from the head-on collisions of equal-mass, non-spinning black holes
    with different values of $\lambda$.
    Bottom: The flux of energy radiated away in the scalar field for the same cases. The different
    cases have been scaled to the highest value of $\lambda$ assuming $\lambda^2$ scaling.
\label{fig:bhbh_rad_comp}
}
\end{figure}

In Fig.~\ref{fig:bhbh_rad_comp}, we show the radiation from the black hole mergers.
Increasing $\lambda$ to larger values slightly decreases the merger time, and
increases the gravitational radiation.  More pronounced is the effect this has
on the scalar radiation, which roughly scales as $\lambda^2$, though on top of
this, some additional nonlinear enhancement is evident for large values.  For
$\lambda/m^2 \gtrsim 0.1$, the scalar field luminosity is comparable to the
gravitational wave luminosity for this configuration.

\begin{figure}
\begin{center}
    \includegraphics[width=\columnwidth,draft=false]{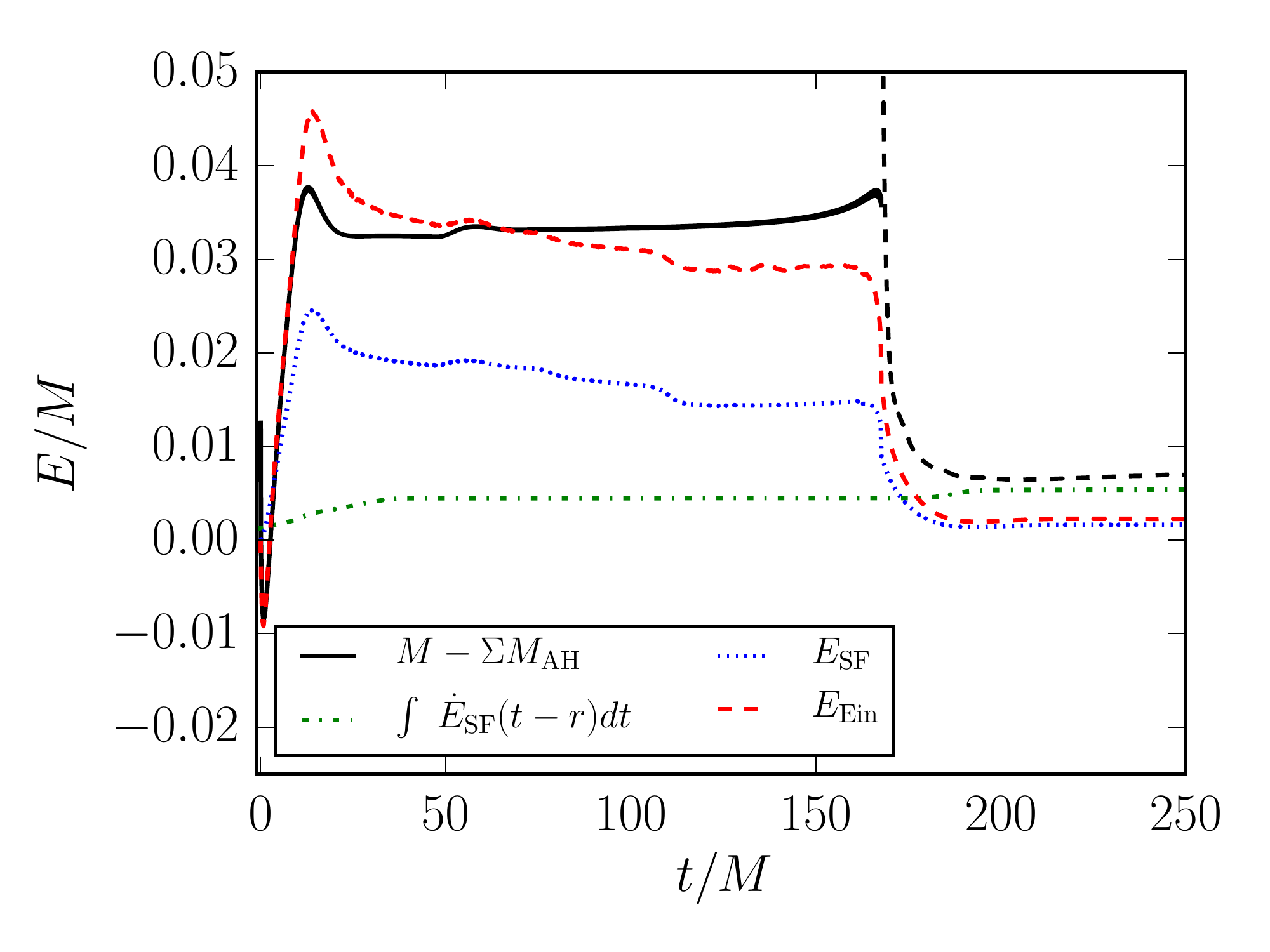}
\end{center}
\caption{
    Various measures of energy as a function of time for a head-on collisions
of equal-mass, non-spinning black holes with $\lambda/m^2=0.15$. We show the difference
of the total mass from the sum of the mass of the apparent horizons, the scalar
field energy radiated away (at $r=50M$), the integrated energy in the canonical scalar field
component $E^{\rm SF}$, and calculated from the Einstein tensor $E^{\rm Ein}$.
We note that the last two quantities are gauge dependent except when the spacetime is 
stationary, which approximately holds at late times.
\label{fig:bhbh_eall}
}
\end{figure}

Even though these black hole merger spacetimes are far from being stationary, except at
late times after the final remnant has settled down, it is still instructive to
study an approximate measure of how energy is distributed as a function of time.
In Fig.~\ref{fig:bhbh_eall}, we show this for a stronger coupling case with
$\lambda/m^2=0.15$. We can see that as the black holes form scalar hair
and as their mass decreases,
there is a comparable increase in the effective energy calculated
from the Einstein tensor $E^{\rm Ein}$, with roughly half of this being
attributable to the canonical scalar field energy $E^{\rm SF}$.  After the
formation of a common horizon, these quantities rapidly decrease.  For this
case, most of energy that does not end up in the final black hole is actually radiated
away as scalar radiation (dotted green curve in Fig.~\ref{fig:bhbh_eall}).
In fact, the initial scalar hair growth
of the individual black holes produces stronger
radiation than the merger.

\begin{figure}
\begin{center}
\includegraphics[width=\columnwidth,draft=false]{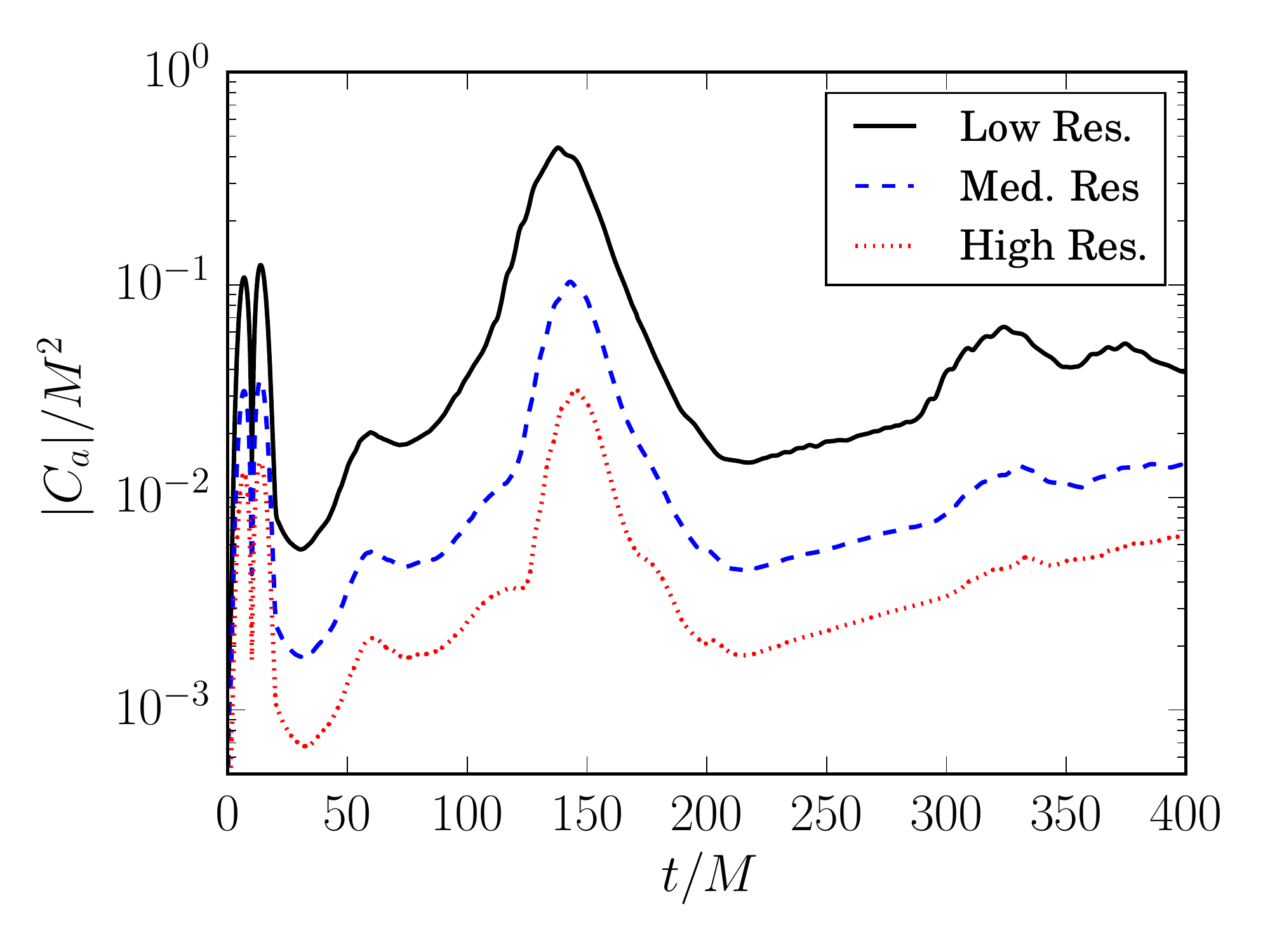}
\end{center}
\caption{
    Integrated norm of the constraint violation (Eq.~\ref{eq:mh_condition}) as
    a function of time (in units of total mass) for an equal-mass binary black hole
    merger with $\lambda/m^2=0.15$ at three resolutions.  The medium and high
    resolutions have $1.5$ and $2\times$ the linear resolution of the low
    resolution simulation.
\label{fig:axisym_cnst_cnv}
}
\end{figure}

We also show the integrated norm of the constraint violation
(Eq.~\ref{eq:mh_condition}) for $\lambda/m^2=0.15$ and several resolutions in
Fig.~\ref{fig:axisym_cnst_cnv}, demonstrating that this quantity is converging
to zero at the expected rate.  Here the lowest resolution has a grid spacing of
$dx\approx 0.02 M$ on the finest level, and the highest resolution is twice as
high.

\begin{figure}
\begin{center}
\includegraphics[width=\columnwidth,draft=false]{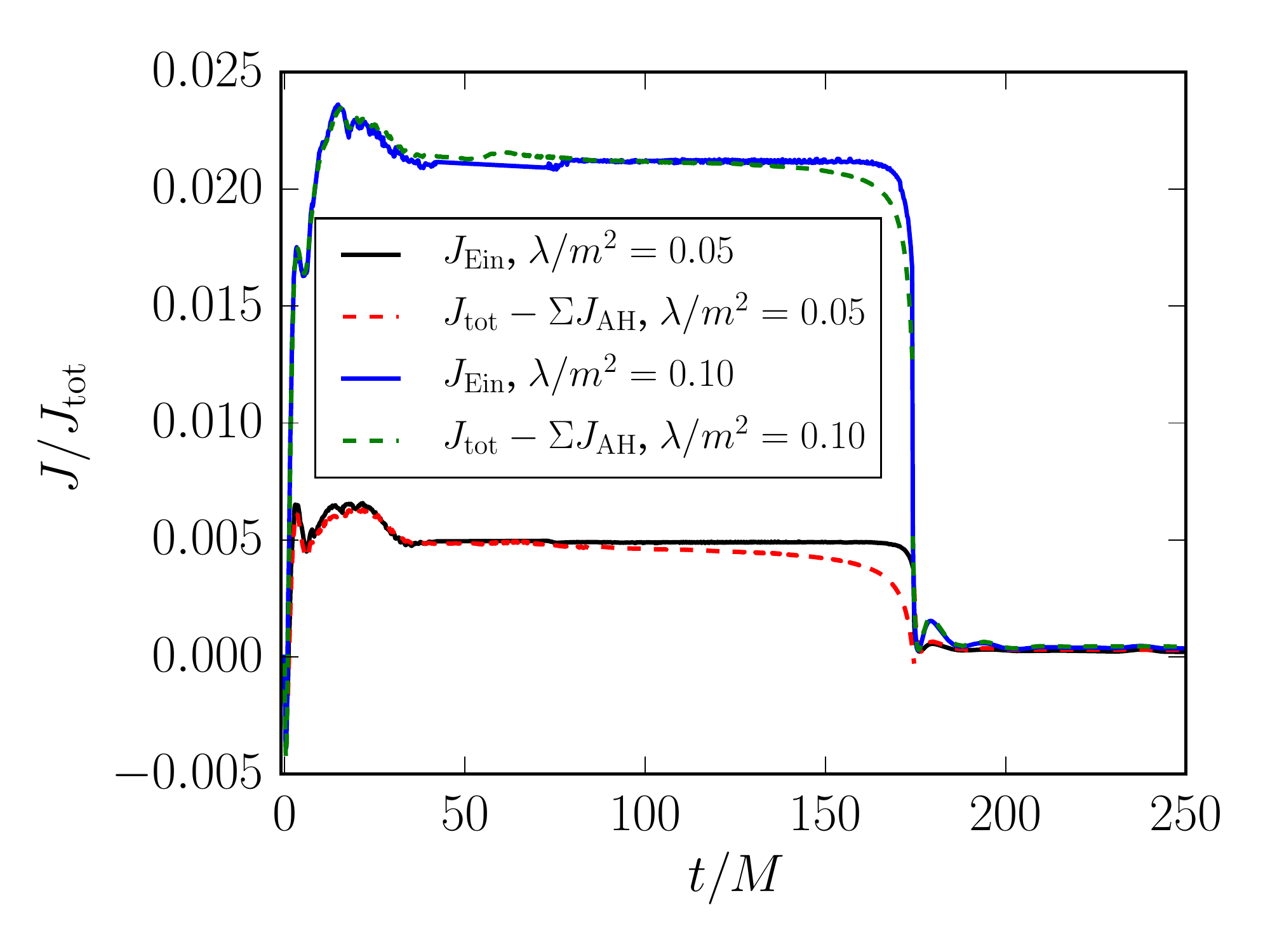}
\end{center}
\caption{
    Angular momentum as a function of time for a head-on collisions of
    equal-mass black holes with aligned $a=0.6$ spins and $\lambda/m^2=0.05$ and 0.1.
    We show the difference of the total angular momentum from the sum of the
    angular momentum of the apparent horizons, and the integrated angular
    momentum calculated from the Einstein tensor $J_{\rm Ein}$.
\label{fig:bhbh_jall}
}
\end{figure}

To probe the effect of angular momentum, we also study mergers of spinning black holes.
We consider two axisymmetric configurations where the magnitude of the
dimensionless black hole spin is $|a|=0.6$: one where the spins are aligned, and one
where they are anti-aligned.  In Fig.~\ref{fig:bhbh_jall}, we show how the
angular momentum evolves in the aligned cases.
Initially, as the black holes grow scalar hair,
angular momentum moves from the black hole horizons to the scalar clouds.
As the black holes merge,
most of this angular momentum goes back into the final black hole.

\begin{figure}
\begin{center}
    \includegraphics[width=\columnwidth,draft=false]{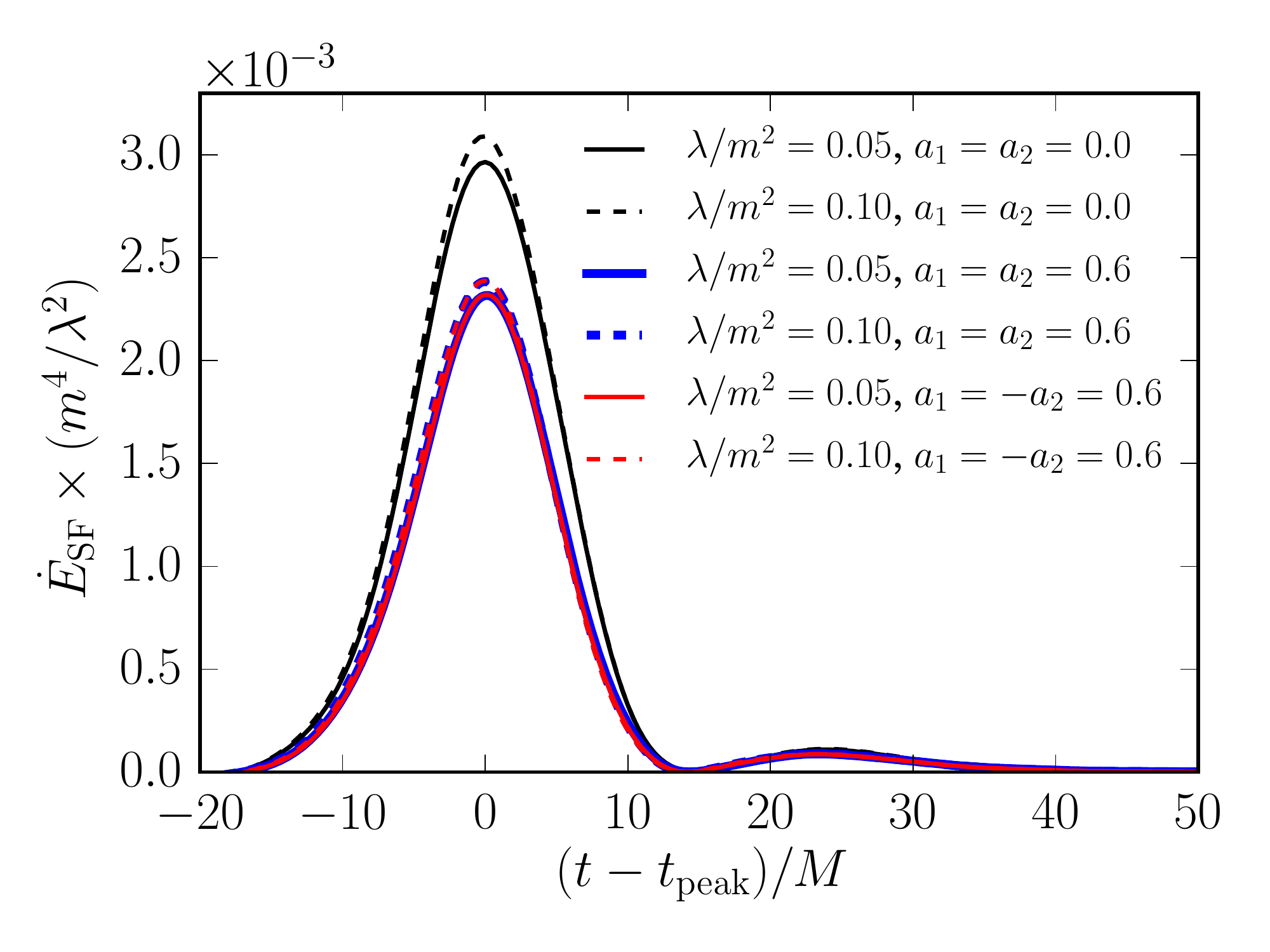}
    \includegraphics[width=\columnwidth,draft=false]{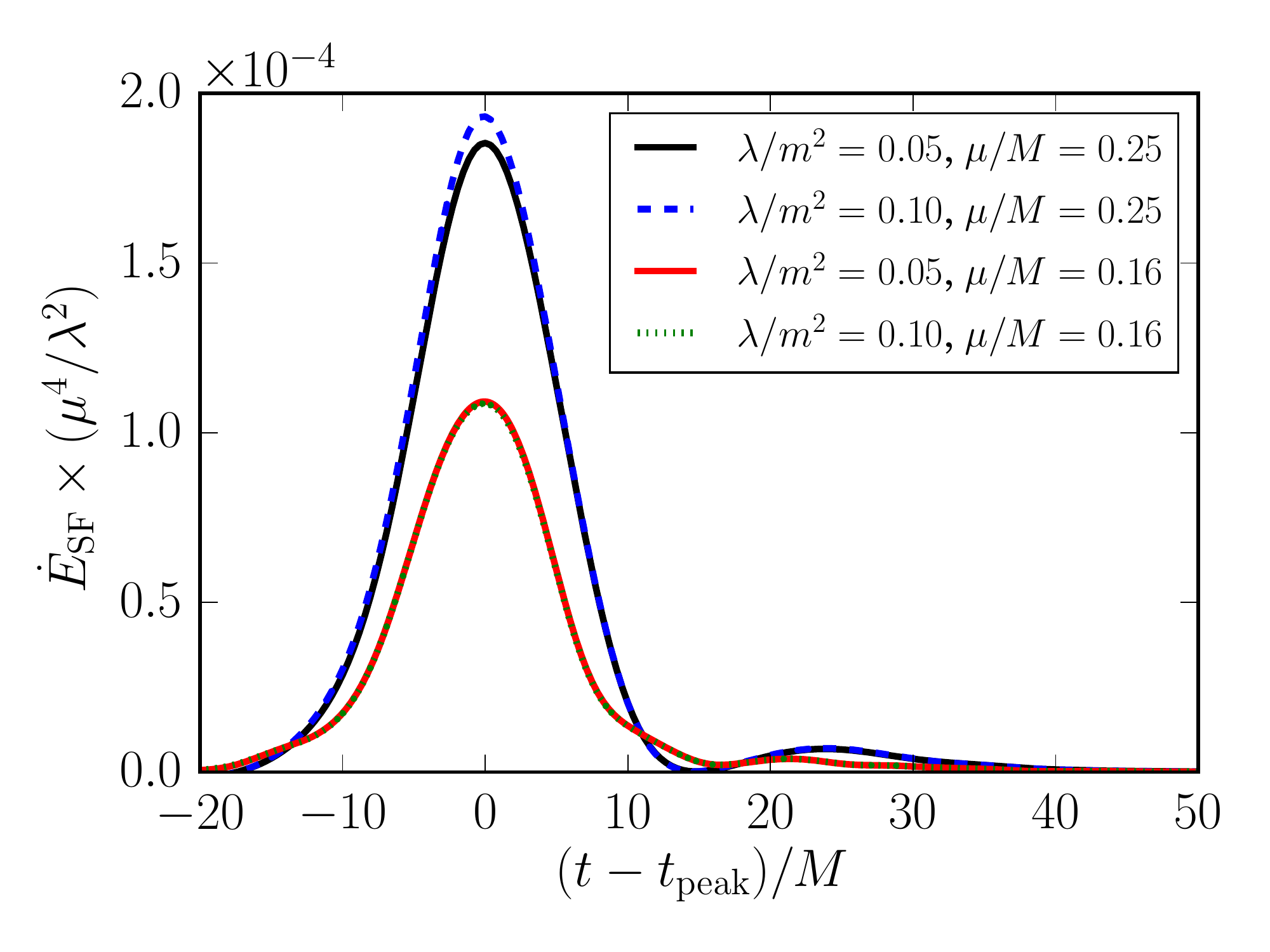}
\end{center}
\caption{
    The scalar luminosity in the wavezone ($r/M=100$) for various head-on black hole mergers.
    Top: A comparison of different values of black hole spin and ESGB coupling for equal mass mergers.
    The luminosity has been scaled assuming a $\lambda^2$ dependence.
    Bottom: A comparison of different values of mass-ratio and coupling for non-spinning mergers.
    The luminosity has been scaled by $\lambda^2/\mu^4$, where $\mu$ is the
    reduced mass of the binary.
\label{fig:bhbh_sf_flux_all}
}
\end{figure}

The scalar radiation produced by the spinning black hole mergers is slightly smaller
compared to the non-spinning cases, as shown in the top panel of
Fig.~\ref{fig:bhbh_sf_flux_all}.  We also find that the differences between the
aligned and anti-aligned spins is negligible for these cases.  We note that in
axisymmetry, the scalar field radiation does not carry angular momentum.

Finally, we consider a 4:1 mass-ratio merger of non-spinning black holes.
In this configuration,
the smaller black hole will have more scalar hair than the larger
one, which tends to suppress non-linear effects in the coupling due to the
merger.  In the bottom panel of Fig.~\ref{fig:bhbh_sf_flux_all}, we show the
scalar radiation from two cases with $\lambda/m^2=0.05$ and 0.1. Compared to
the equal-mass cases, the luminosity is smaller by roughly a factor of
$\sim12$. After rescaling by $\lambda^2$, the $\lambda/m^2=0.05$ and 0.1 cases
are indistinguishable for a 4:1 mass ratio.
\subsection{3D results: scalar hair formation about a boosted,
   spinning black hole
   }
   We next discuss results for spinning, boosted black hole initial data.
To consider a fully 3D example, we choose the initial spin axis
and boost axis to be unaligned---e.g. the initial spin of the black hole
is in the $z$ direction, and the boost is in the $y$ direction.
As discussed in Sec.~\ref{sec:initial_data}, our initial data for
the scalar field is: $\phi=\partial_0\phi=0$, so that we initially start
out with a boosted, spinning black hole (in harmonic coordinates),
which subsequently forms a scalar cloud.
Therefore, unlike in the vacuum Einstein equations case, the black hole boost
is more than just a coordinate transformation.
The main result of this section is that the boosted,
spinning black hole spontaneously form scalar hair and that we obtain stable,
convergent evolution. 

\begin{figure*}
\begin{center}
    \includegraphics[width=\columnwidth,draft=false]{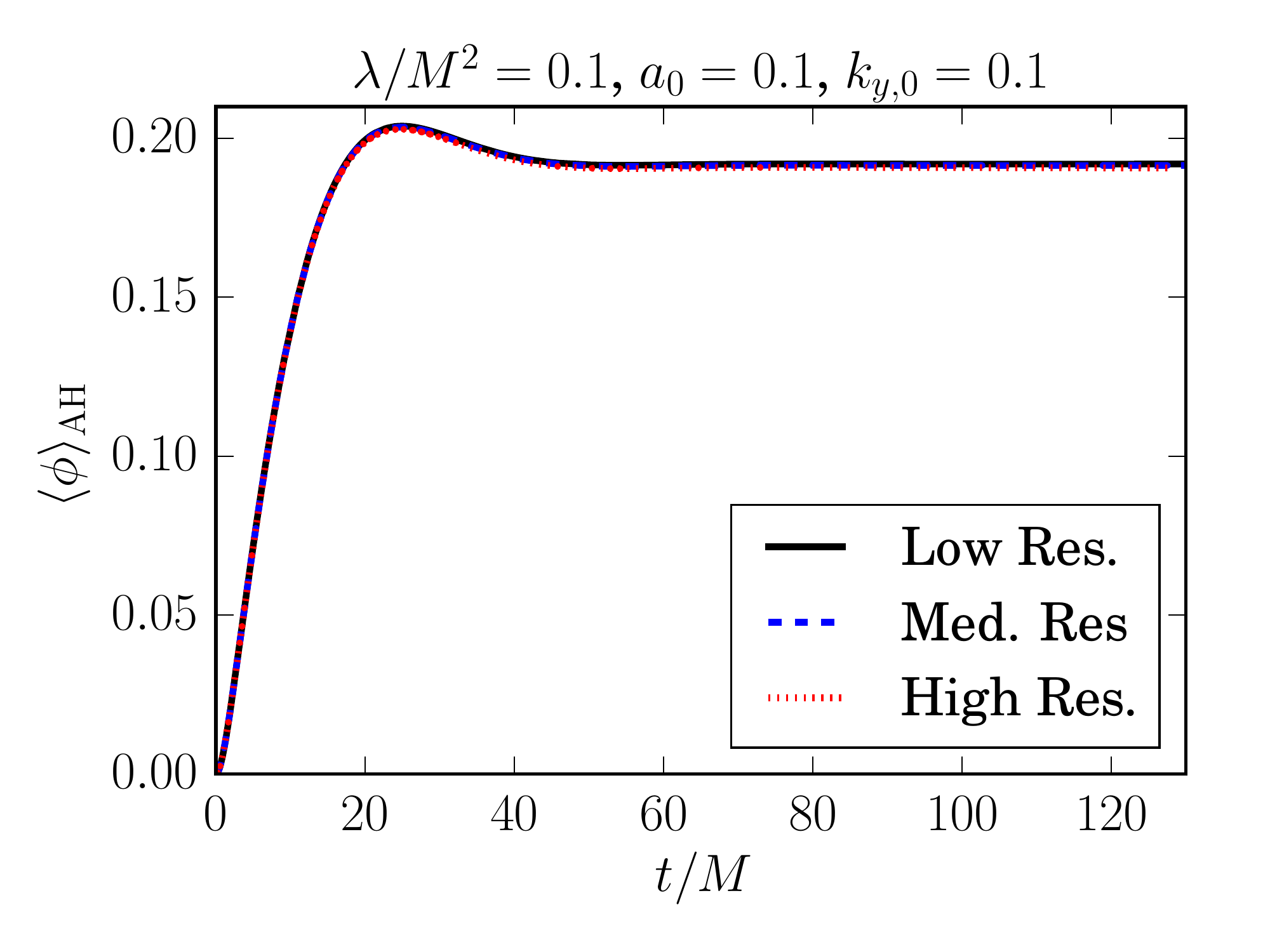}
    \includegraphics[width=\columnwidth,draft=false]{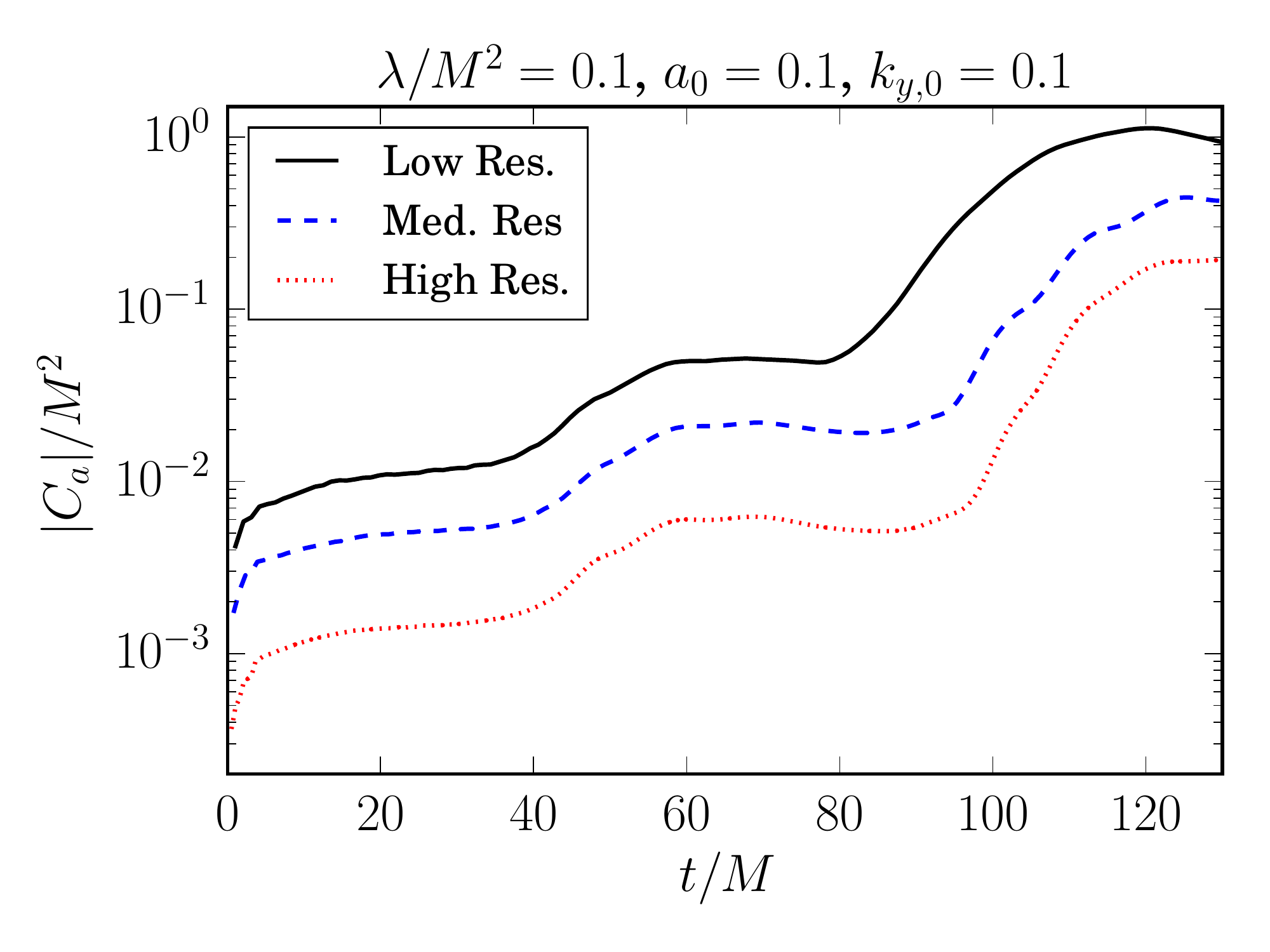}
\end{center}
\caption{
  Convergence study of the horizon-averaged scalar field
$\left<\phi\right>_{AH}$ (left) and constraint violation $|C_a|$ (normalized
by the initial black hole mass $M$; right), for a fully 3D case 
with a black hole with initial boost $k_{y,0}=0.1$, dimensionless spin $0.1$,
and coupling $\lambda/M^2=0.1$. 
We find that the constraint violation converges at third order.
The transient growth at $t\sim100\ M$ in this quantity is due to the 
scalar radiation from initial scalar hair growth hitting the outermost
mesh refinement level,
which leads to some spurious reflection (which converges away). 
The medium and high
resolutions have $1.5$ and $2\times$ the linear resolution of the low
resolution simulation.
}
\label{fig:3d_run}
\end{figure*}

   We show one example case in this section: a Kerr black hole
with initial dimensionless spin $a=0.4$, and with a boost $k_{y,0}=0.1$
(i.e. at $10\%$ the speed of light) orthogonal to the initial spin axis.
(We found similar results for other cases with higher spins and lower values of coupling,
e.g. $a=0.2$ and $\lambda/M^2=0.05$.)
In Fig.~\ref{fig:3d_run}, we show a convergence study of the constraint
violation $|C^a|$ and find third order convergence, with no sign of
resolution dependent growth.
Our results for the boosted, spinning black hole are
qualitatively similar to our simulations of spinning black holes
in axisymmetry: a scalar field grows and then settles down to an equilibrium
configuration around the black hole, emitting a burst of scalar radiation in the process. 

\subsection{3D results: binary black hole inspiral and merger}
\label{sec:3dbinaryinspiral}
Finally, we consider the inspiral and merger of a binary black hole without continuous symmetries.
Here we just present results for one case consisting of an equal-mass, non-spinning binary that undergoes $\sim 3$
orbits before merging, and for a relatively small value of the Gauss-Bonnet coupling
$\lambda/m^2=0.01$ (where again, $m$ refers to the mass of one constituent of the binary),
where nonlinear effects are small.
As part of the process for constructing initial data, we evolve the binary
black hole data obtained from solving the constraint equations for $\sim 50M$ (where $M$ is the ADM mass)
just using the Einstein equations, to reduce the gauge dynamics and spurious high frequency
gravitational wave content. We then use this as initial data 
for the evolution with the full ESGB equations.

\begin{figure}
\begin{center}
    \includegraphics[width=\columnwidth,draft=false]{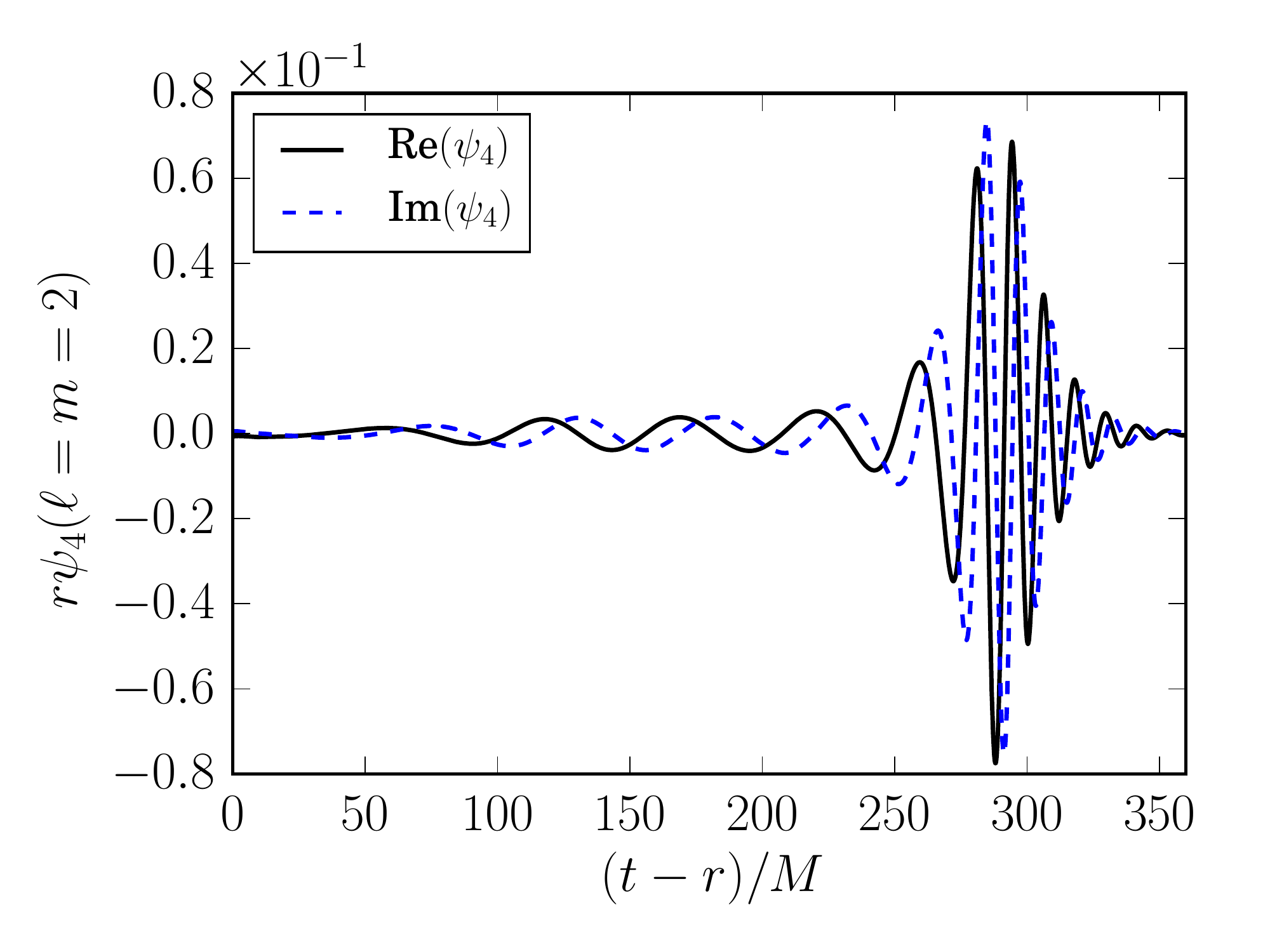}
    \includegraphics[width=\columnwidth,draft=false]{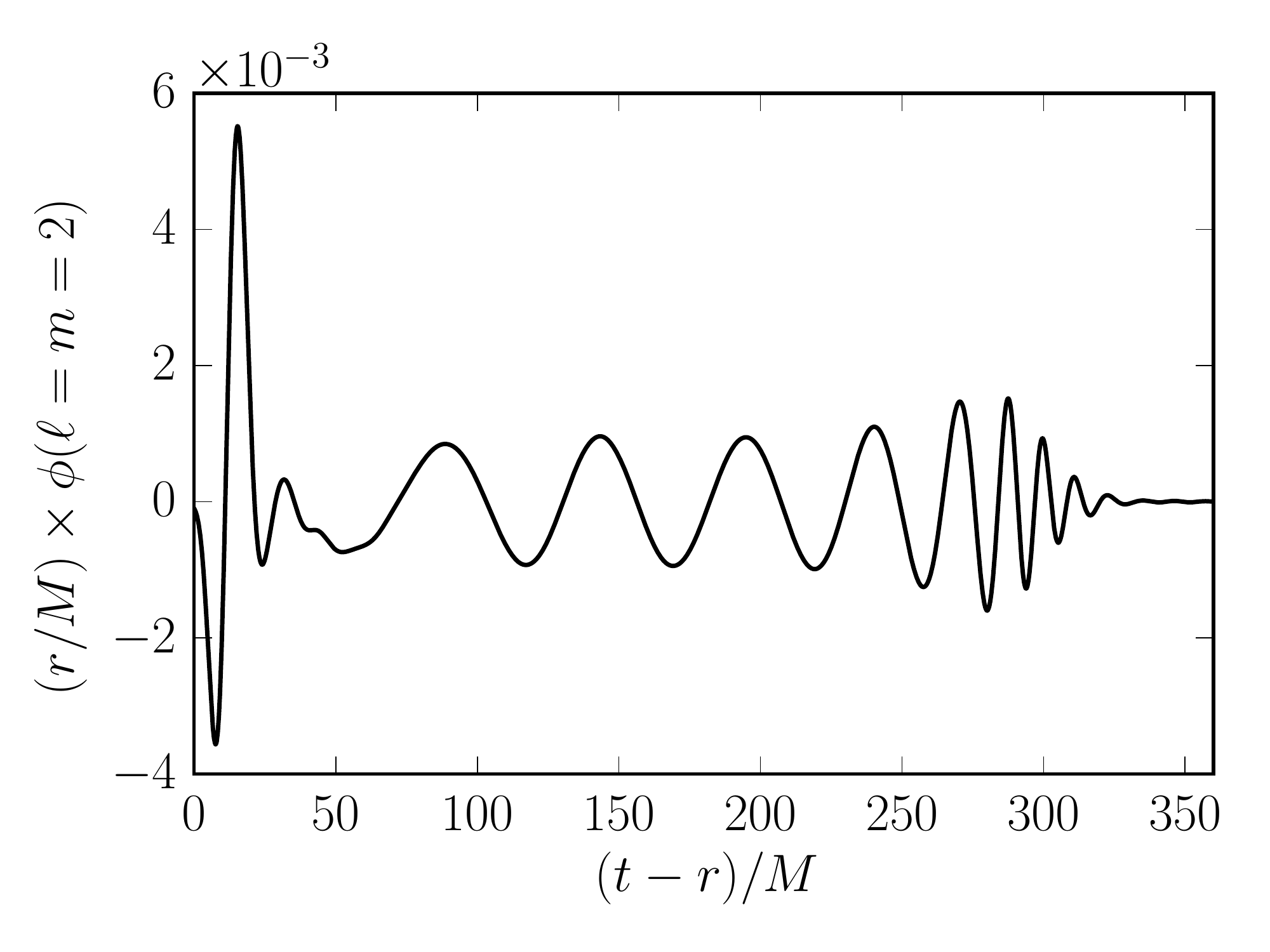}
\end{center}
\caption{
    The radiation from the inspiral and merger of an equal mass binary black hole with $\lambda/m^2=0.01$.
    Top: The real and imaginary components of the $\ell=m=2$ spin $-2$ spherical harmonic of the Newman-Penrose scalar,
    which encodes the gravitational waves.
    Bottom: The $\ell=m=2$ component of the scalar field in the wavezone.
The burst at early times comes from the growth of the scalar hair about
the individual black holes  of the initially vacuum binary.
\label{fig:qc_bhbh_rad}
}
\end{figure}

As was found for the head-on black hole merger,
the black holes rapidly form scalar hair,
after which the scalar
field around the individual horizons is essentially constant,
with a small uptick in the
last stages of the merger that is cutoff by the appearance of the common horizon.
We show the gravitational and scalar radiation for this system in Fig.~\ref{fig:qc_bhbh_rad}.
As was seen in the head-on black hole mergers, the burst of scalar radiation from the initial growth of the scalar hair about the individual black holes
of the binary is actually larger than for the merger.
Following this initial transient, the scalar radiation
tracks the inspiral of the binary evident in the gravitational waves. 
As expected from the results in Sec.~\ref{sec:axisym_bhbh}, at this small value of the coupling
the scalar radiation is much smaller than the gravitational radiation. 
Ignoring the initial burst, which is just an artifact of our initial conditions,
$\approx 3\times10^{-6} M$ is emitted in scalar radiation during the last few orbits, mostly at merger 
(compared to a few percent $M$ emitted in gravitational waves).

\begin{figure}
\begin{center}
    \includegraphics[width=\columnwidth,draft=false]{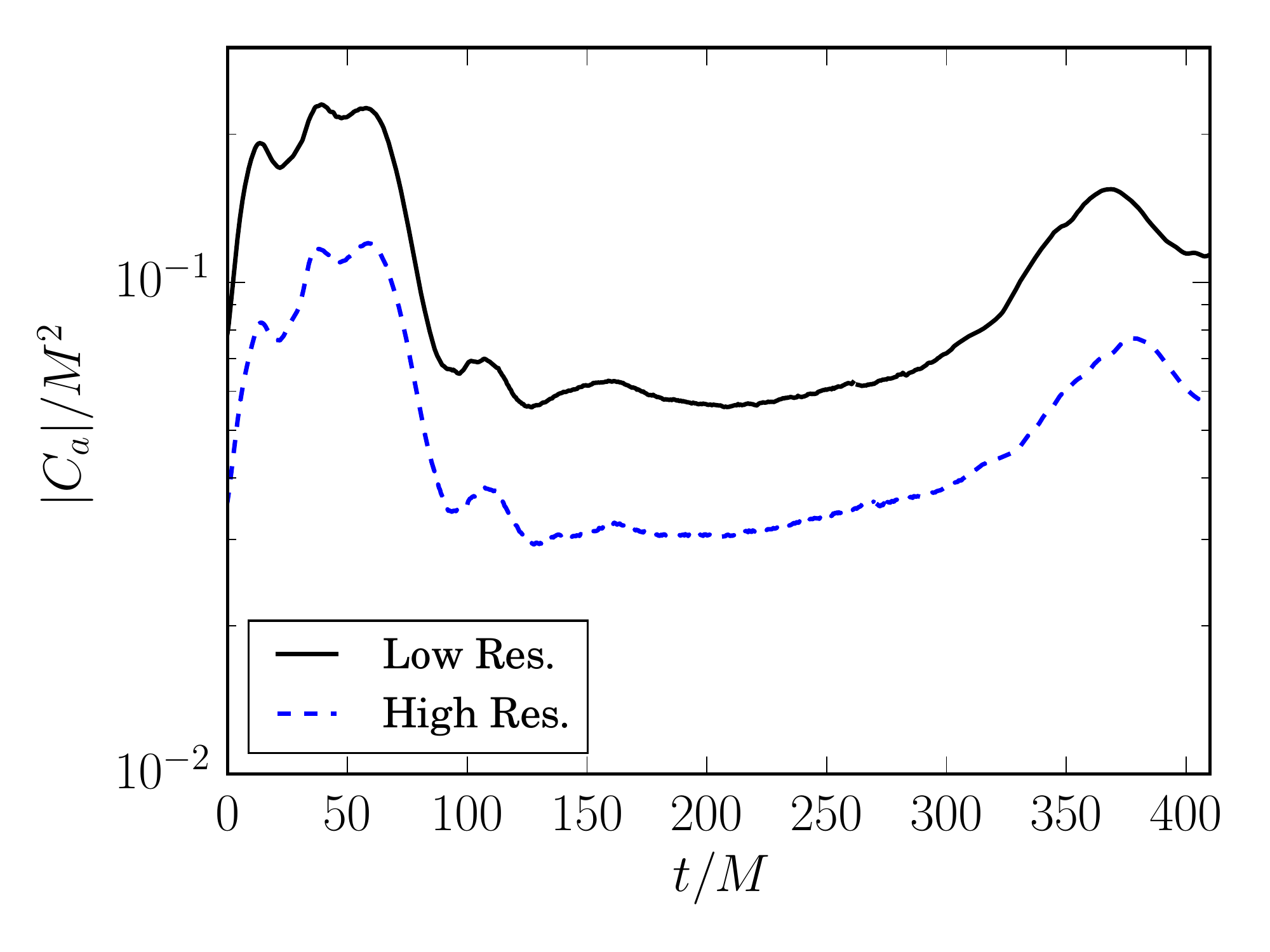}
\end{center}
\caption{
   The integrated norm of the MGH constraint violation $|C_a|$ for the inspiral
and merger of an equal-mass binary black hole with $\lambda/m^2=0.01$.  The high
resolution case has $4/3\times$ the resolution of the low resolution, and the
convergence is consistent with third order.
\label{fig:qc_bhbh_cnst}
}
\end{figure}

We perform this calculation at two different resolutions where the lower resolution 
has a grid spacing of $dx/M\approx 0.025$  on the finest mesh refinement level,
and the higher resolution has $4/3\times$ this resolution. We show the time dependence of the integrated MGH
constraint in Fig.~\ref{fig:qc_bhbh_cnst}. This is consistent with third-order convergence, and shows no
sign of resolution dependent growth.

We present these results primarily to illustrate that these methods work for the binary inspiral
problem, at least at sufficiently modest values of the modified gravity coupling.
We leave the exploration of higher values of the coupling and different
binary configurations, as well as additional resolution studies, to future work.

\section{Discussion and Conclusion}
\label{sec:conclusion}
   In this article we present numerical solutions of
dynamical spacetimes in ESGB gravity without any particular symmetry
restrictions or approximations. 
We evolve the shift-symmetric ESGB EOMs using the MGH formulation, 
for which the theory has a well-posed initial value
problem (at least at weak coupling) \cite{Kovacs:2020pns,Kovacs:2020ywu}.
We are able to solve for the dynamics of single and binary
(scalar hairy) black hole spacetimes in this theory,
including cases with spinning black holes and binaries with unequal mass-ratios,
and in the regime where the formation of scalar hair changes
the black hole mass at the level of a few percent,
and the scalar radiation becomes comparable to the gravitational radiation.
Given the novelty of the modified harmonic formulation, and the dearth
of results on the nonlinear dynamics of Horndeski theories
in regimes of physical interest, there are many avenues for future research.
In this section we outline a few such directions.

Here, we presented one case consisting of a few orbits and merger
of a quasi-circular binary black hole at a relatively modest value
of the Gauss-Bonnet coupling, 
to demonstrate that our methods work for such configurations. 
However, based on the results from head-on collisions, 
higher values of coupling, where nonlinear effects due to the ESGB
terms are more important, should be tractable.
In future work, we will explore the parameter space
of binary black hole mergers more thoroughly,
in order to make a better connection
to gravitational wave observations,
which can be used test such modifications to 
GR. Here, we have focused on shift-symmetric ESGB, but our methods
should be generally applicable both to ESGB with other couplings,
and Horndeski gravity theories in general,
potentially allowing binary black hole mergers in all these
cases to be explored.
Direct simulations of the full EOMs could also be compared
to various approximate treatments of these theories, including
the order reduction approach~\cite{Okounkova:2017yby},
or modified forms of the EOMs that are designed
to improve the hyperbolicity~\cite{Cayuso:2017iqc},
in order to quantify the errors coming from secular
or non-perturbative effects in a binary inspiral (such a comparison
for a toy scalar-field problem was carried out in Ref.~\cite{Allwright:2018rut}). 
This would help determine the best methods to use for theories where the short
wavelength behavior is not known, and there is not a well-posed initial value 
problem.

   In tackling the above,
another future research direction is to better understand the 
robustness of the MGH formulation, both of the Einstein equations,
and of the Horndeski gravity theories, under different gauge choices. 
In this new formulation, one can freely choose not only the source functions,
but also the two auxiliary metrics $\tilde{g}^{ab}$ and $\hat{g}^{ab}$
which determine the light cone of the gauge
and constraint propagating modes.  Our particular choice of the auxiliary
metrics (Eq.~\ref{eq:relation_gtilde_ghat_g}) was guided mostly by convenience,
and it would be interesting to look for improved choices of auxiliary metrics
that could, for example allow us to better evolve black hole spacetimes with
larger ESGB coupling.
Potential future directions include: considering different ratios
of the parameters $\tilde{A}$ and $\hat{A}$, and considering
different ansaetze for the auxiliary metrics $\tilde{g}^{ab}$
and $\hat{g}^{ab}$ (for example, it would be considering auxiliary metrics
where $\tilde{A}$ and $\hat{A}$ are functions of the spacetime geometry;
for more discussion see \cite{Kovacs:2020ywu}).

   Another research direction is to develop robust initial
data solution methods for the Horndeski theories.
This will be necessary for numerically
constructing initial data for, e.g. inhomogeneous cosmological solutions
to Horndeski theories, or binary initial data that does not have a strong
initial transient due to, e.g. initial black hole scalar hair formation.
The first step in this direction would be to formulate the Horndeski constraints
as elliptic equations, for example using a conformal thin-sandwich
type approach~\cite{kovacsrealltalk}.

   Finally, given the number of Horndeski theories that have been invoked in
both the early and late universe
(e.g.
\cite{Kobayashi:2011nu,Ijjas:2016vtq,Creminelli:2010ba,Clifton:2011jh,Kobayashi:2019hrl}),
a natural direction for future research
is to consider cosmological solutions to Horndeski theories.
Given the failure of GR coupled to ordinary matter
to resolve the initial cosmological singularity \cite{hawking1973large},
it would be interesting to determine if \emph{any} classical field
theory that had well-posed evolution could resolve this issue in
a mathematically satisfactory way, while also obeying current observational
and experimental constraints. Potential candidate theories that have
been proposed (e.g. bouncing universes \cite{Ijjas:2016vtq}, or
``genesis'' \cite{Creminelli:2010ba,Ageeva:2020buc}), fall under
the Horndeski class of theories, and thus should be amenable to being
solved using a MGH formulation.
 
\section*{Acknowledgements}
   We are grateful to Aron Kovacs and Harvey Reall for several helpful
discussions about modified harmonic gauge, the hyperbolicity of
Horndeski gravity theories, and 4$\partial$ST gravity.
W.E. acknowledges support from an NSERC Discovery grant.
This research was
supported in part by Perimeter Institute for Theoretical Physics.  Research at
Perimeter Institute is supported by the Government of Canada through the
Department of Innovation, Science and Economic Development Canada and by the
Province of Ontario through the Ministry of Research, Innovation and Science.
This research was enabled in part by support provided by SciNet
(www.scinethpc.ca/) and Compute Canada (www.computecanada.ca).
Some of the simulations presented in this article were performed on
computational resources managed and supported by Princeton Research Computing,
a consortium of groups including the Princeton Institute for
Computational Science and Engineering (PICSciE)
and the Office of Information Technology's High Performance
Computing Center and Visualization Laboratory at Princeton University.
\onecolumngrid
\appendix
\section{Derivation of evolution matrix for
4$\partial$ST gravity in a modified
harmonic formulation}
\label{appen:derivation_evolution_matrix}
   For completeness, and for reference, here we
show our derivation of the components
of the evolution matrix, Eq.~\ref{eq:time_evo_system},
for the EOMs of 4$\partial$ST gravity, 
Eqs.~\ref{eq:eom_edgb_scalar} and \ref{eq:eom_edgb_tensor}.

We find it convenient to split our calculation into several steps:
first we rewrite the equations for the Einstein-minimally coupled
scalar field contributions
to the EOMs, then for the contributions 
that involve $\alpha$, and then, finally, for the contributions
that involve $\beta$.
\subsubsection{Terms:
   Einstein, minimally coupled scalar field, and
   constraint damping
}
   We first consider the Einstein-modified harmonic contribution
to the tensor EOMs 
\begin{align}
   R_{ab}
-  \left(
      \hat{P}_c{}^d{}_{ab}
   -  \frac{1}{2}g_{ab}\hat{P}_c{}^d
   \right)
   \nabla_dC^c
-  \frac{1}{2}\kappa\left(
      n_aC_b
   +  n_bC_a
   -  \left(1+\rho\right)n_cC^cg_{ab}
   \right)
   =
   8\pi\left(
      T_{ab}
   -  \frac{1}{2}g_{ab}T
   \right)
   .
\end{align}
   It is straightforward to see that
\begin{align}
   B_{ab}{}=0
   .
\end{align}

    We next consider
\begin{align}
    \nabla_dC^c
    =&
    \nabla_d\left(H^c+\tilde{g}^{ef}\Gamma_{ef}^c\right)
    \nonumber\\
    =&
    \frac{1}{2}\tilde{g}^{ef}g^{cg}
    \left(
            \partial_d\partial_eg_{gf}
    +	\partial_d\partial_fg_{ge}
    -	\partial_d\partial_gg_{ef}
    \right)
    \nonumber\\&
+	\tilde{g}^{ef}\partial_dg^{cg}\Gamma_{gef}
+	\partial_d\tilde{g}^{ef}\Gamma^c_{ef}
+	\partial_dH^c
   \nonumber\\&
+  \Gamma^c_{dg}\left(H^g+\tilde{g}^{ef}\Gamma^g_{ef}\right)
    .
\end{align}
   Using
\begin{subequations}
\begin{align}
    R_{ab}
    =&
    \partial_c\Gamma^c_{ab}
-	\partial_a\Gamma^c_{cb}
+	\Gamma^c_{dc}\Gamma^d_{ab}
-	\Gamma^c_{da}\Gamma^d_{cb}
    \nonumber\\
    =&
-	\frac{1}{2}g^{cd}\left(
            \partial_c\partial_dg_{ab}
    -	\partial_c\partial_bg_{ad}
    -	\partial_a\partial_dg_{bc}
    +	\partial_a\partial_bg_{cd}
    \right)
    \nonumber\\&
+	\partial_cg^{cd}\Gamma_{dab}
-	\partial_ag^{cd}\Gamma_{dcb}
+	\Gamma^c_{dc}\Gamma^d_{ab}
-	\Gamma^c_{da}\Gamma^d_{cb}
    ,\\
    \partial_cg^{cd}\Gamma_{dab}
+	\Gamma^c_{dc}\Gamma^d_{ab}
    =&
-	\Gamma_d\Gamma^d_{ab}
    ,\\
    \partial_ag^{cd}\Gamma_{dcb}
    =&
    \frac{1}{2}\partial_{a}g^{cd}\partial_{b}g_{dc}
    \nonumber\\
    =&
    \frac{1}{4}\partial_{a}g^{cd}\partial_{b}g_{dc}
+	\frac{1}{4}\partial_{b}g^{cd}\partial_{a}g_{dc}
    ,
\end{align}
\end{subequations}
    we then have
\begin{align}
\label{eq:intermediate_eqn_einstein}
-  \frac{1}{2}A_{ab}{}^{cdef}\partial_c\partial_dg_{ef}
-  \frac{1}{4}\partial_{a}g^{cd}\partial_{b}g_{cd}
-  \frac{1}{4}\partial_{b}g^{cd}\partial_{a}g_{cd}
-  \Gamma_d\Gamma^d_{ab}
-  \Gamma^c_{da}\Gamma^d_{cb}
   \nonumber\\
-  \left(
      \hat{P}_c{}^d{}_{ab}
   -  \frac{1}{2}g_{ab}\hat{P}_c{}^d
   \right)
   \left(
      \partial_dH^c
   +  \Gamma_{gef}\tilde{g}^{ef}\partial_dg^{cg}
   +  \Gamma^c_{ef}\partial_d\tilde{g}^{ef}
   +  \Gamma^c_{dg}\left(
         H^g+\tilde{g}^{ef}\Gamma^g_{ef}
      \right)
   \right)
   \nonumber\\
-  \frac{1}{2}\kappa\left(
      n_aC_b
   +  n_bC_a
   -  \left(1+\rho\right)n_cC^cg_{ab}
   \right)
   \nonumber\\
   =
   8\pi\left(
      T_{ab}
   -  \frac{1}{2}T g_{ab}
   \right)
    ,
\end{align}
    where
\begin{align}
    A_{ab}{}^{cdef}
    \equiv&
    \delta_a^e\delta_b^fg^{cd}
-	\delta_a^f\delta_b^dg^{ce}
-	\delta_a^c\delta_b^fg^{de}
+	\delta_a^c\delta_b^dg^{ef}
    \nonumber\\&
+	2\left(\hat{P}^{ec}{}_{ab}-\frac{1}{2}g_{ab}\hat{P}^{ec}\right)
    \tilde{g}^{df}
-	\left(\hat{P}^{dc}{}_{ab}-\frac{1}{2}g_{ab}\hat{P}^{dc}\right)
    \tilde{g}^{ef}
    .
\end{align}
    Note that we can interchange $c\leftrightarrow d$
and $e\leftrightarrow f$, as partial derivatives commute and
$g_{ef}$ is symmetric. We use this fact below to simplify some of
the expressions. 
    To see the structure of the principal symbol in more detail
we expand out $\hat{P}^{cd}{}_{ab}$
\begin{align}
\label{eqn:Ptr}
    \hat{P}^{dc}{}_{ab}
-	\frac{1}{2}g_{ab}\hat{P}^{dc}
    =&
    \frac{1}{2}\left(
            \delta_a^d\hat{g}_b{}^c
    +	\delta_b^d\hat{g}_a{}^c
    -	g^{cd}\hat{g}_{ab}
    -	g_{ab}\left(
                    \hat{g}^{dc}
            -	\frac{1}{2}g^{cd}\hat{g}
            \right)
    \right)
    .
\end{align}
    We then have
\begin{align}
\label{eqn:Asimplified}
    A_{ab}{}^{cdef}
    =&
    \delta_a^e\delta_b^fg^{cd}
-	\left(
            \delta_a^f\delta_b^dg^{ce}
    -	\delta_a^f\hat{g}_b{}^d\tilde{g}^{ce}
    \right)
-	\left(
            \delta_a^c\delta_b^fg^{de}
    -	\delta_b^f\hat{g}_a{}^c\tilde{g}^{de}
    \right)
+	\left(
            \delta_a^c\delta_b^dg^{ef}
    -	\delta_{(a}^d\hat{g}^c_{b)}\tilde{g}^{ef}
    \right)
    \nonumber\\&
-	\left(
            g^{ce}\hat{g}_{ab}
    +	\hat{g}^{ce}g_{ab}
    -	\frac{1}{2}g_{ab}g^{ce}\hat{g}
    \right)\tilde{g}^{df}
+	\frac{1}{2}\left(
            g^{cd}\hat{g}_{ab}
    +	g_{ab}\hat{g}^{cd}
    -	\frac{1}{2}g_{ab}g^{cd}\hat{g}
    \right)\tilde{g}^{ef}
    .
\end{align}
   From Eq.~\ref{eq:intermediate_eqn_einstein},
we can read off $A_{ab}{}^{cd}$ and $F^{(g)}_{ab}$: 
\begin{align}
   A_{ab}{}^{cd}
   =&
   A_{ab}{}^{00cd}
   ,\\
   F^{(g)}_{ab}
   =&
    A_{ab}{}^{\alpha\beta ef}\partial_{\alpha}\partial_{\beta}g_{ef}
+  2A_{ab}{}^{(\alpha 0) ef}\partial_{\alpha}\partial_{0}g_{ef}
   \nonumber\\&
-  \frac{1}{4}\partial_ag^{cd}\partial_bg_{cd}
-  \frac{1}{4}\partial_bg^{cd}\partial_ag_{cd}
-  \Gamma_d\Gamma^d_{ab}
-  \Gamma^c_{da}\Gamma^d_{cb}
   \nonumber\\&
-  \left(
      \hat{P}_c{}^d{}_{ab}
   -  \frac{1}{2}g_{ab}\hat{P}_c{}^d
   \right)
   \left(
      \partial_dH^c
   +  \Gamma_{gef}\tilde{g}^{ef}\partial_dg^{cg}
   +  \Gamma^c_{ef}\partial_d\tilde{g}^{ef}
   +  \Gamma^c_{dg}\left(H^g+\tilde{g}^{ef}\Gamma^g_{ef}\right)
   \right)
   \nonumber\\&
-  \frac{1}{2}\kappa\left(
      n_aC_b
   +  n_bC_a
   -  \left(1+\rho\right)n_cC^cg_{ab}
   \right)
\nonumber\\&
-  8\pi\left(
      T_{ab}
   -  \frac{1}{2}T g_{ab}
   \right)
   ,
\end{align}

   The contribution of the scalar field is 
\begin{align}
   F_{ab}^{(g)}
   =&
-  \nabla_a\phi\nabla_b\phi
-  V(\phi)g_{ab}
   ,\\
   D
   =&
   g^{00}
   ,\\
   F^{(\phi)}
   =&
    g^{\alpha\beta}\partial_{\alpha}\partial_{\beta}\phi
+  2g^{\alpha 0}\partial_{\alpha}\partial_{0}\phi
-  g^{ab}\Gamma_{ab}^c\partial_c\phi
   .
\end{align}
   In Appendix \ref{eq:reduction_eom_Einstein}, we explicitly show how
Eq.~\ref{eq:intermediate_eqn_einstein} reduces
in the special case of $\hat{g}^{ab}=\tilde{g}^{ab}=g^{ab}$ to the
Einstein equations in the generalized harmonic formulation with constraint
damping.
\subsubsection{Term: $\alpha$}
   We next consider the terms that involve $\alpha$.
There are no second derivative terms on the metric, so we have
\begin{align}
    A_{ab}{}^{cd}
    =&
    0
    \\
    B_{ab}
    =&
    0
    ,\\
    C^{cd}
    =&
    0
    .
\end{align}
    The nonzero terms are
\begin{align}
   D
   =&
   2\alpha\left(\phi\right)\left(
      Xg^{00}
   -  \nabla^0\phi\nabla^0\phi
   \right)
   ,\\
   F^{(g)}_{ab}
   =&
-  2\alpha\left(\phi\right)X\nabla_a\phi\nabla_b\phi
-  \alpha\left(\phi\right)X^2g_{ab}
   \\
   F^{(\phi)}
   =&
   4\alpha\left(\phi\right)\left(
      Xg^{\alpha0}
   -  \nabla^{\alpha}\phi\nabla^{0}\phi
   \right)
   \partial_{\alpha}\partial_{0}\phi
   \nonumber\\&
+  2\alpha\left(\phi\right)\left(
      Xg^{\alpha\beta}
   -  \nabla^{\alpha}\phi\nabla^{\beta}\phi
   \right)
   \partial_{\alpha}\partial_{\beta}\phi
   \nonumber\\&
-  2\alpha\left(\phi\right)\left(X g^{cd} - \nabla^c\phi\nabla^d\phi\right)
   \Gamma^c_{ab}\partial_c\phi
-  3\alpha^{\prime}\left(\phi\right)X^2
    .
\end{align}
\subsubsection{Term: $\beta$}
   Finally we consider the terms that involve the Gauss-Bonnet scalar.
Due to the length of the necessary algebraic manipulations,
we write things out in stages. First we expand
\begin{subequations}
\begin{align}
    R^{ij}{}_{ef}
    =&
    g^{jk}R^i{}_{kef}
    ,\nonumber\\
    =&
    g^{jk}\left(
        \partial_e\Gamma^i_{kf}
    -   \partial_f\Gamma^i_{ke}
    +   \Gamma^i_{el}\Gamma^l_{kf}
    -   \Gamma^i_{fl}\Gamma^l_{ke}
    \right)
    \nonumber\\
    =&
    \frac{1}{2}g^{jk}g^{im}\left(
        \partial_e\partial_kg_{mf}
    +   \partial_e\partial_fg_{mk}
    -   \partial_e\partial_mg_{kf}
    -   \partial_f\partial_kg_{me}
    -   \partial_f\partial_eg_{mk}
    +   \partial_f\partial_mg_{ke}
    \right)
    \nonumber\\&
    +   g^{jk}\left(
        \partial_eg^{im}\Gamma_{mkf}
    -   \partial_fg^{im}\Gamma_{mke}
    +   \Gamma^i_{el}\Gamma^l_{kf}
    -   \Gamma^i_{fl}\Gamma^l_{ke}
    \right)
    \\
    \nabla^g\nabla_c\beta\left(\phi\right)
    =&
    g^{gl}\left(
        \partial_l\partial_c\beta\left(\phi\right)
    -   \Gamma^m_{lc}\partial_m\beta\left(\phi\right)
    \right)
    \nonumber\\
    =&
    g^{gl}\left(
        \beta^{\prime}\left(\phi\right)
        \left[
            \partial_l\partial_c\phi
         -   \Gamma^m_{lc}\partial_m\phi
         \right]
    +   \beta^{\prime\prime}\left(\phi\right)
        \partial_l\phi\partial_c\phi
    \right)
    .
\end{align}
\end{subequations}
    We find that
\begin{align}
\label{eq:beta_expanded tensor_eom}
   2\delta^{efcd}_{ijg(a}g_{b)d}R^{ij}{}_{ef}
      \nabla^g\nabla_c\beta(\phi)
-  \delta^{efc}_{ijg}R^{ij}{}_{ef}
      \nabla^g\nabla_c\beta(\phi) g_{ab}
   = \nonumber\\
    g^{jk}g^{gl}
    \Delta^{efc}_{ijgab}
    \left(
        g^{im}\partial_e\partial_kg_{mf}
    +   \partial_eg^{im}\Gamma_{mkf}
    +   \Gamma^i_{em}\Gamma^m_{kf}
    \right)
    \nonumber\\\times
    \left(
        \beta^{\prime}\left(\phi\right)
        \left[
            \partial_l\partial_c\phi
         -  \Gamma^m_{lc}\partial_m\phi
         \right]
    +   \beta^{\prime\prime}\left(\phi\right)
        \partial_l\phi\partial_c\phi
    \right)
    , 
\end{align}
   where we have defined the tensor
\begin{align}
   \Delta^{efc}_{ijgab}
   \equiv
   2\left(
      2\delta^{efcd}_{ijg(a}g_{b)d}
   -  \delta^{efc}_{ijg}g_{ab}
   \right)
   ,
\end{align}
   which is antisymmetric on the top three and first bottom three
indices, and symmetric for the rightmost two bottom indices.

   It turns out that while the EOMs for 4$\partial$ST
gravity are fully nonlinear, they are linear with respect to
repeated derivatives; e.g. there are no terms like
$(\partial_0^2g_{ab})(\partial_0^2\phi)$
or
$(\partial_0^2g_{ab})^2$ in the equations of motion
(see Appendix \ref{appendix:properties_principal_symbol} for
an explicit calculation; we note that this property holds more
generally for all Horndeski gravity theories
\cite{Papallo:2017qvl,Papallo:2017ddx}).
Thus, there is no ambiguity in computing terms like $A_{ab}{}^{cd}$
and $B_{ab}$.

   From Eq.~\ref{eq:beta_expanded tensor_eom} we have
\begin{align}
   A_{ab}{}^{cd}
   =&
   g^{j0}\Delta^{0d\gamma}_{ijgab}g^{lg} 
   g^{ic}
   \left(
      \beta^{\prime}\left(\phi\right)
      \left[
         \partial_l\partial_{\gamma}\phi
      -  \Gamma^m_{l\gamma}\partial_m\phi
      \right]
   +  \beta^{\prime\prime}\left(\phi\right)\partial_l\phi\partial_{\gamma}\phi
   \right)
   ,\\
   B_{ab}
   =&
   g^{g0}g^{jk}\Delta^{\gamma \rho 0}_{ijgab}
   \left(
      g^{im}\partial_{\gamma}\partial_{k}g_{m\rho}
   +  \partial_{\gamma}g^{im}\Gamma_{mk\rho}
   +  \Gamma^i_{\gamma m}\Gamma^m_{k\rho}
   \right)
   \beta^{\prime}\left(\phi\right)
   \\
   F_{ab}
   =&
   g^{jk}g^{gl}\Delta^{efc}_{ijgab}
   \left(
      g^{im}\partial_e\partial_kg_{mf}
   +  \partial_eg^{im}\Gamma_{mkf}
   +  \Gamma^i_{em}\Gamma^m_{kf}
   \right)
   \nonumber\\
   &\times\left(
      \beta^{\prime}\left(\phi\right)
      \left[
         \partial_l\partial_c\phi
      -  \Gamma^m_{lc}\partial_m\phi
      \right]
   +  \beta^{\prime\prime}\left(\phi\right)
      \partial_l\phi\partial_c\phi
   \right)
   \nonumber\\&
   -A_{ab}{}^{cd}\partial_0^2g_{cd}
   -B_{ab}\partial_0^2\phi
   .
\end{align}

    We next look at the scalar field EOM. 
The Gauss-Bonnet scalar is
\begin{align}
    \mathcal{G}
    \equiv&
    \frac{1}{4}\delta^{pqrs}_{ghij}R^{gh}{}_{pq}R^{ij}{}_{rs}
    \nonumber \\
    =&
    \delta^{pqrs}_{ghij}
    \left(
        g^{hk}g^{gm}\partial_p\partial_kg_{mq}
    +   g^{hk}\partial_pg^{gm}\Gamma_{mkq}
    +   g^{hk}\Gamma^g_{pm}\Gamma^m_{kq}
    \right)
    \nonumber\\ &\times
    \left(
        g^{jv}g^{iw}\partial_r\partial_vg_{ws}
    +   g^{jv}\partial_rg^{iw}\Gamma_{wvs}
    +   g^{jv}\Gamma^i_{rw}\Gamma^w_{vs}
    \right)
    .
\end{align}
    We then have
\begin{align}
    D
    =
    0
    ,
\end{align}
   and
\begin{align}
    C^{cd}
    =&
    2\beta^{\prime}\left(\phi\right)\delta^{\alpha\beta0d}_{ghij}
    \left(
        g^{hk}g^{gm}\partial_{\alpha}\partial_kg_{m\beta}
    +   g^{hk}\partial_{\alpha}g^{gm}\Gamma_{mk\beta}
    +   g^{hk}\Gamma^g_{\alpha m}\Gamma^m_{k\beta}
    \right)
    g^{j0}g^{ic}
    ,\\
    F^{(\phi)}
    =&
   \beta^{\prime}\left(\phi\right)\mathcal{G}
-  C^{cd}\partial_0^2g_{cd}
   .
\end{align}
\section{Properties of the principal part of ESGB gravity}
\label{appendix:properties_principal_symbol} 
  The evolution equations for ESGB gravity,
Eqs.~\ref{eq:eom_edgb_scalar} and \ref{eq:eom_edgb_tensor},
form a fully nonlinear system of partial differential equations.
It turns out though (and this is a general property of the EOMs
of Horndeski gravity theories) that
the EOMs do not contain terms with repeated derivatives, e.g. terms like 
$(\partial_c\partial_cg_{ab})^2$.
In this section, we review the derivation of this fact.
We consider the equations in the form of
Eq.~\ref{eq:time_evo_system}.
\begin{enumerate}[(i)]
\item $A_{ab}{}^{cdef}$: The only second order derivative
term is from $\partial_l\partial_p\phi$. Consider then
$c=d=l=p=Z$. We then have $\delta^{ZqZf}_{ghij}=0$
and $\delta^{ZZf}_{gij}=0$.
\item $B_{ab}{}^{cd}$: The only nonzero second order
term is $\partial_e\partial_kg_{mf}$. Set then $c=d=e=k=Z$.
We then have $\delta^{ZZef}_{ghij}=0$
and $\delta^{ZZf}_{gij}=0$.
\item $C^{cdef}$: The only term second order in derivatives is
$\partial_p\partial_kg_{mq}$. Set $c=d=p=k=Z$. We then have
$\delta^{ZqZf}_{ghij}=0$.
\item $D^{ab}=0$ for all $a,b$.
\end{enumerate}
\section{Reduction of modified harmonic formulation to generalized
harmonic formulation}
\label{eq:reduction_eom_Einstein}
   For reference, here we demonstrate
how the EOMs for GR in the MGH 
formulation reduce to those of a generalized harmonic formulation
for the special choice that $\hat{g}^{ab}=\tilde{g}^{ab}=g^{ab}$. 
Beginning with Eqs.~\ref{eqn:Ptr} and \ref{eqn:Asimplified}, 
  we set $\tilde{g}^{ab}=\hat{g}^{ab}=g^{ab}$ to obtain
\begin{align}
    A_{ab}{}^{cdef}
    =&
    \delta_a^e\delta_b^fg^{cd}
    ,\\
   \hat{P}_c{}^d{}_{ab}
-  \frac{1}{2}g_{ab}\hat{P}_c{}^d
   =&
   g_{c(a}\delta^d_{b)}
   ,
\end{align}
    so that the Einstein equations given by Eq.~\ref{eq:intermediate_eqn_einstein} become
\begin{align}
-	\frac{1}{2}g^{cd}\partial_c\partial_c\partial_dg_{ab}
-	\frac{1}{2}\partial_{(a}g^{cd}\partial_{b)}g_{dc}
-	\Gamma^c_{da}\Gamma^d_{cb}
+	H_d\Gamma^d_{ab}
    \nonumber\\
    g_{c(a}\delta^d_{b)}\left(
            \partial_dH^c
    +	\Gamma_{gef}g^{ef}\partial_dg^{cg}
    +	\Gamma^c_{ef}\partial_dg^{ef}
   +  \Gamma^c_{dg}\left(
         H^g+g^{ef}\Gamma^g_{ef}
   \right)
    \right)
    \nonumber\\
-	\frac{1}{2}\kappa\left(
            n_aC_b
    +	n_bC_a
    -	\left(1+\rho\right)n_cC^cg_{ab}
    \right)
    \nonumber\\
    =
    8\pi\left(T_{ab}-\frac{1}{2}T g_{ab}\right)
    .
\end{align}
    Simplifying, we obtain the Einstein equations
in a generalized harmonic formulation with constraint
damping terms (e.g. Ref.~\cite{Pretorius:2004jg})
\begin{align}
    -\frac{1}{2}g^{cd}\partial_c\partial_dg_{ab}
-	\partial_cg_{d(a}\partial_{b)}g^{cd}
-	\nabla_{(a}H_{b)}
+	H_c\Gamma^c_{ab}
-	\Gamma^c_{da}\Gamma^d_{cb}
    \nonumber\\
-	\frac{1}{2}\kappa\left(
            n_aC_b
    +	n_bC_a
    -	\left(1+\rho\right)n_cC^cg_{ab}
    \right)
    \nonumber\\
    =
    8\pi\left(
      T_{ab}
   -  \frac{1}{2}Tg_{ab}
   \right)
    .
\end{align}

\bibliography{../mod_grav}

\end{document}